\documentclass[showpacs,preprintnumbers,amsmath,amssymb,pre]{revtex4}

\usepackage{graphics}
\usepackage{ulem}
\usepackage{amsmath}
\usepackage{graphicx,color}
\usepackage[latin1]{inputenc}
\usepackage[T1]{fontenc}
\usepackage{ae,aecompl}

\begin{document}

\title{Order in  glassy  systems}
\author{Jorge Kurchan}

\affiliation{PMMH-ESPCI UMR 7636, 10 rue Vauquelin, 75231 Paris, France\\
  Institut Henri Poincar\'e, 11 rue Pierre et Marie Curie, 75231 Paris, France}
\author{Dov Levine }

\affiliation{Department of Physics, Technion, Haifa 32000, Israel}


\begin{abstract}
{A directly measurable correlation length may be defined for  systems  having a two-step relaxation, 
based on the geometric properties of 
 density  profile that remains after  averaging out the fast motion.
    We argue that the length diverges if
 and when the slow timescale diverges, whatever the microscopic mechanism at the origin of  the slowing down.
 Measuring the length amounts to determining explicitly  the complexity
 from the observed particle configurations. One may compute in the same way the Renyi complexities $K_q$, their relative behavior 
 for different $q$  
   characterizes the mechanism  underlying the  transition.
In particular, the 'Random First Order' scenario predicts that  in the glass phase $K_q=0$ for $q>x$, and $K_q>0$ for $q<x$, 
with $x$ the Parisi 
parameter.  The hypothesis of a  nonequilibrium effective temperature may also be directly tested directly from configurations. }
\end{abstract}
\maketitle

\section{Introduction}

A solid is an arrangement of particles with permanent density modulations.  While commonplace, it is nevertheless remarkable that a system of soft particles may condense into a crystalline state, since due to thermal activation, no particle is strictly confined to a specific region of space. This notwithstanding, permanent density modulations persist forever in an infinite sample.  The reason for this is that in order for spontaneous thermal fluctuations to
completely erase the density modulations in the solid phase,  rearrangements involving a divergent number of
particles are required, and this in turn requires infinite free-energies, and as such is an infinitely rare event.

Glasses also appear solid, in the sense that they have long-lived density modulations, although it is still not known whether these are truly permanent as for crystals.  Similarly, relaxation in some supercooled liquids occurs on a timescale that appears to diverge faster than Arrhenius as the temperature is lowered.  Both these features of slow relaxation, if they hold rigorously even for a system of soft particles, 
may only be explained if there is an ever increasing number of  particles having correlated evolutions, implying that there
is a growing coherence length \cite{Montanari} as a function of decreasing temperature.
The central puzzle of glasses is that there does not appear to be any recognizable form of spatial order - on the contrary, the configurations seem definitely liquid-like, so it is not clear from whence they derive their solidity. 

There are several possible resolutions to this: the simplest is to conclude that glasses are not truly solids. That is,
their relaxation would  become purely Arrhenius at sufficiently low temperatures, as is expected for a system comprised of independent parts. In this case, an amorphous  solid would strictly speaking exist only in the trivial case of hard particles immobilized by contact with others, for which all activation is absent.
Another possibility is that (at least some) glasses indeed have a truly super-Arrhenius behavior as $T \rightarrow 0$ or even as $T \rightarrow T_K$, for some finite temperature $T_K$, termed the ``Kauzmann temperature'' \cite{Kauzmann}. 
This assumption is implicit in several theories of fragile glasses.   For example, density functional theory expresses the liquid properties in terms of a free energy functional of the the density, and exhibits, for low temperatures and high pressures,
 not only periodic 'crystalline' configurations, but also non-periodic, amorphous states \cite{DFT}.  A related, more complete version of this is  
 the `Random First Order' scenario~\cite{RFOT, Cavagna}, which posits  an `ideal glass' state 
 to which the system would equilibrate given an infinitely slow annealing from high temperature to 
 $T < T_{K}$.   Such an ideal state would presumably be characterized by permanent
non-periodic spatial density-modulations.  In order for the solids postulated in these theories to exist in finite dimensions beyond the mean-field approximation, they require coherence  at diverging lengths~\cite{foot000}.  This leads to a central question: Can this cooperativity be observed directly from configurations? 

In this paper we shall discuss a family of lengths~\cite{JD} that, we argue, should diverge in any system having a timescale that diverges at some temperature (or equivalently, shows a super-Arrhenius behavior as $T \rightarrow 0$).  The lengths are geometric in nature, in the sense that they do not require information on either microscopic dynamics, the history of the sample, or even whether the system is in or out of equilibrium.  The various scenarios for the glass transition are distinguished by the relative manner in which the different lengths behave. 

The reasoning proceeds as follows.  
First, we recognize that order concerns the time-averaged density, which is insensitive to fast excitations such as vibrations, particle-vacancy excitations, or isolated spin flips.  In fact, this is true as well for crystals, which are only truly periodic once such excitations are averaged out~\cite{foot00}.
 In a non-crystalline  system, these average density profiles (which we shall term `{\it states}') are not periodic, so a multitude of possibilities for the density profiles are obtained simply by translations. This leads naturally  to the question: How many possible
 { {states}} are there?  The logarithm of this number is  the  {\it complexity} (we reserve the word 'entropy' for quantities based on counting individual configurations, and 'complexity' for those associated to classes of configurations).  We note that care must be taken here, because density distributions are continuous objects, so that counting them is much like
 counting the number of trajectories of a dynamical system. We shall exploit this analogy.
 
 At this point we make the following observation: if we wish to concentrate on a truly solid phase, in which a density profile lasts forever, the
 number of different possible density profiles cannot be exponential in the volume, or, to be more precise, the complexity
 cannot be extensive. This is because if there are exponentially many metastable states in competition, a simple nucleation argument suggests that they cannot all have infinite lifetime; starting from any particular state,  some other state will nucleate in finite time. (This is of course only true for short range, finite interactions in finite dimensions.) As we shall argue, the reason is that if the number of possible finite-size motifs  the system may nucleate is exponential
 in their volume, the increase in probability of nucleating a new motif overcomes the cost, in terms of probability,  of creating an interface.  

We now consider a sample of very large size ${\cal{V}}$, and `patches' of volume $V \ll {\cal{V}}$ belonging to it. 
The fact that the number of density profiles in a  solid is subexponential  immediately implies that patches of configurations of volume $V$ repeat more often than exponentially { {rarely}} in $V$,
so that their number ${\cal N}(V)$ obeys $\ln ({\cal N}(V)) < O(V)$.
On the contrary, within the liquid phase, or in a defective (nonequilibrium) glass, the system breaks into uncorrelated regions, and we have, for large $V$,  
$\ln ({\cal N}(V)) \sim V K_1 $,
with $K_1$ being the complexity per unit volume in the large $V$ limit ($\lim_{V \rightarrow \infty} K_1^V$), with the statistics taken over patches belonging to the (much larger) volume ${\cal{V}}$.

The fact that patterns of all sizes repeat less often than exponentially, so that there is no { {(extensive)}} complexity, directly implies the existence of a divergent
correlation length.
We may see this most easily as follows. Suppose we have a region of  volume $V$ with a configuration $A$. To what extent does $A$ determine the
configuration { {(say, B)}} of a neighboring region, also  of size $V$?   
The  typical number of repeats of configurations of volume $V$ (or rather, its typical logarithm) is $\ln {\cal{N}}(A) \sim \ln {\cal{N}}(B) \sim - K_1^V$. Clearly, if $A$ and $B$ appeared independently, the repetition probability of the pair $A + B$  
would be
$\ln {\cal{N}}(A + B) = \ln {\cal{N}}(A)+ \ln {\cal{N}}(B) \sim -2VK_1^V$, while in general it will be some number  $\ln {\cal{N}}(A +  B) \sim -VK_1^{2V}$.
The patches will then not be independent if $K_1^{2V}<2K_1^V$, or, in other words, if  the entropy of patches of size $V$ is subextensive. In that case,
the knowledge of the configuration in a volume $V$ gives information on that of an adjoining  patch.
Note that we have assumed here that for large $V$, the most likely patch entropy (based on the logarithm of the frequency of appearance of a patch)
  is, to leading order in $V$, independent of the shape. 
If we have an extensive patch entropy $K_1$, and we assume that for large volumes the system may be considered as composed of a number of independent components
of volume $V_1\sim \ell_1^D$, then ${\cal{N}}\sim e^{V K_1} \sim e^{V/V_1}$, and we identify a correlation volume as $V_1 \sim 1/K_1$. 

\begin{figure}[htbp]
\begin{center}
\includegraphics[width=6cm]{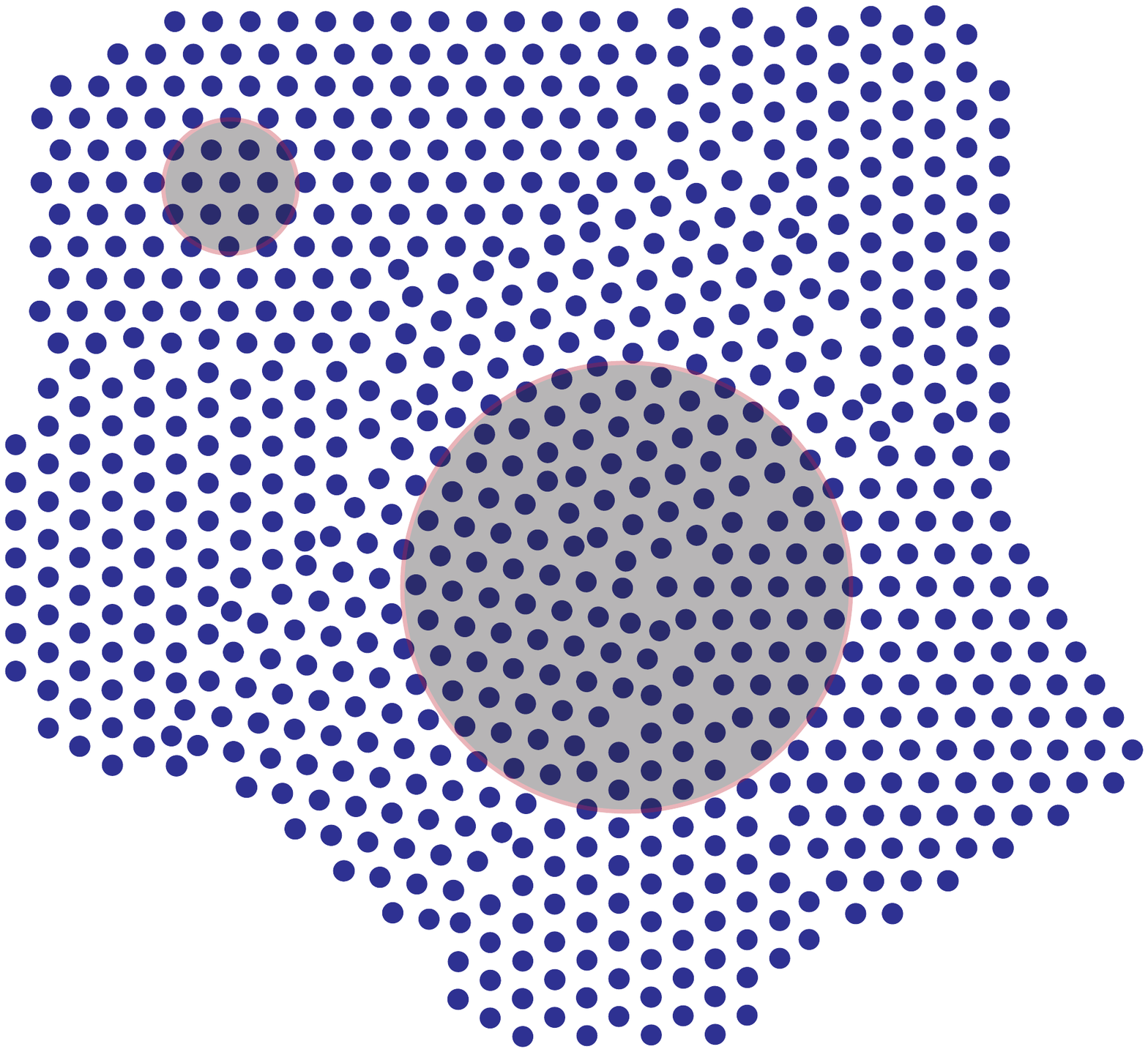}
\end{center}
\caption{ Two patterns, one smaller and one larger than  the crystallite length. 
  The small pattern will recur with high  frequency, but the large
  one will be  rare, because it requires recurrence of the crystallite boundaries within a patch, which is random event. Patches
  that are very large with respect to the crystallite volume occur with multiplicity exponential in the number of defects per volume, of the order of the number of microcrystallites contained. }
\label{mosaic1}
\end{figure}

A system with subextensive  patch entropy $K_1$ 
will then have an infinite length - in this sense, it is {\it ordered}. Now, let us start from such an ordered system, and fracture it into pieces of typical size $V_c$.  Let us now reassemble these pieces at random to form a new system of large size.  In this case,we expect patches smaller than $V_c$ will be found often, and those larger than $V_c$ 
will be exponentially rare, as in Fig. \ref{mosaic1}. This leads us then to the definition of  the following types of length for an amorphous system:
  
  \begin{itemize}
  \item {\bf Complexity length}: $\ell_1= K_1^{-\frac1D}$,  measures the frequency  with which a {\it typical} patch repeats. 
  \item {\bf Renyi lengths}: In order to distinguish different { {glass transition}} scenarios, it is useful to consider patches that repeat more often or less often than is typical. This leads to the definition of lengths  based on the Renyi complexities, or, equivalently,  the {\it large deviation function}  for patch repetition frequencies.
  \item {\bf Crossover length}: patches with volumes smaller than $V \ll V_c$ repeat often, while larger ones $V_c \ll V$ are exponentially rare. The crossover length is $\ell \sim V_c^{\frac{1}{D}}$.

 A related measure is the following:  suppose the complexity of a patch  of volume $V$ is given, with finite $V$ corrections, by:
    $V K_1 -  A(V) K^{(1)}_1+  ...$, where $A(V)$ is subleading in $V$, for example $A(V)\sim V^\nu$. The next to leading term $A(V) K^{(1)}_1$ describes correlations, and is hence referred to as  the 'synergy'  \cite{Bialek}. $K^{(1)}_1$ might diverge at the critical point \cite{foot111}.  If $A(V)$ is not $O(1)$ in $V$,  $\frac{K^{(1)}_1}{K_1}$ provides a  new length.
  \end{itemize}
  
  These  measures may be understood easily in the case of a microcrystalline system, as sketched in Figure \ref{mosaic1}.  However, one of the central points of our proposal is that it allows us to define a correlation length for systems whose ground states are not only non-periodic, but even conventionally thought of as amorphous, in the sense that their scattering patterns have no Bragg peaks.  In what follows, we shall discuss in detail how the counting and identification of patters may be made, as well as the implications to 
current glass scenarios.

\begin{figure}[htbp]
\begin{center}
\includegraphics[width=6cm]{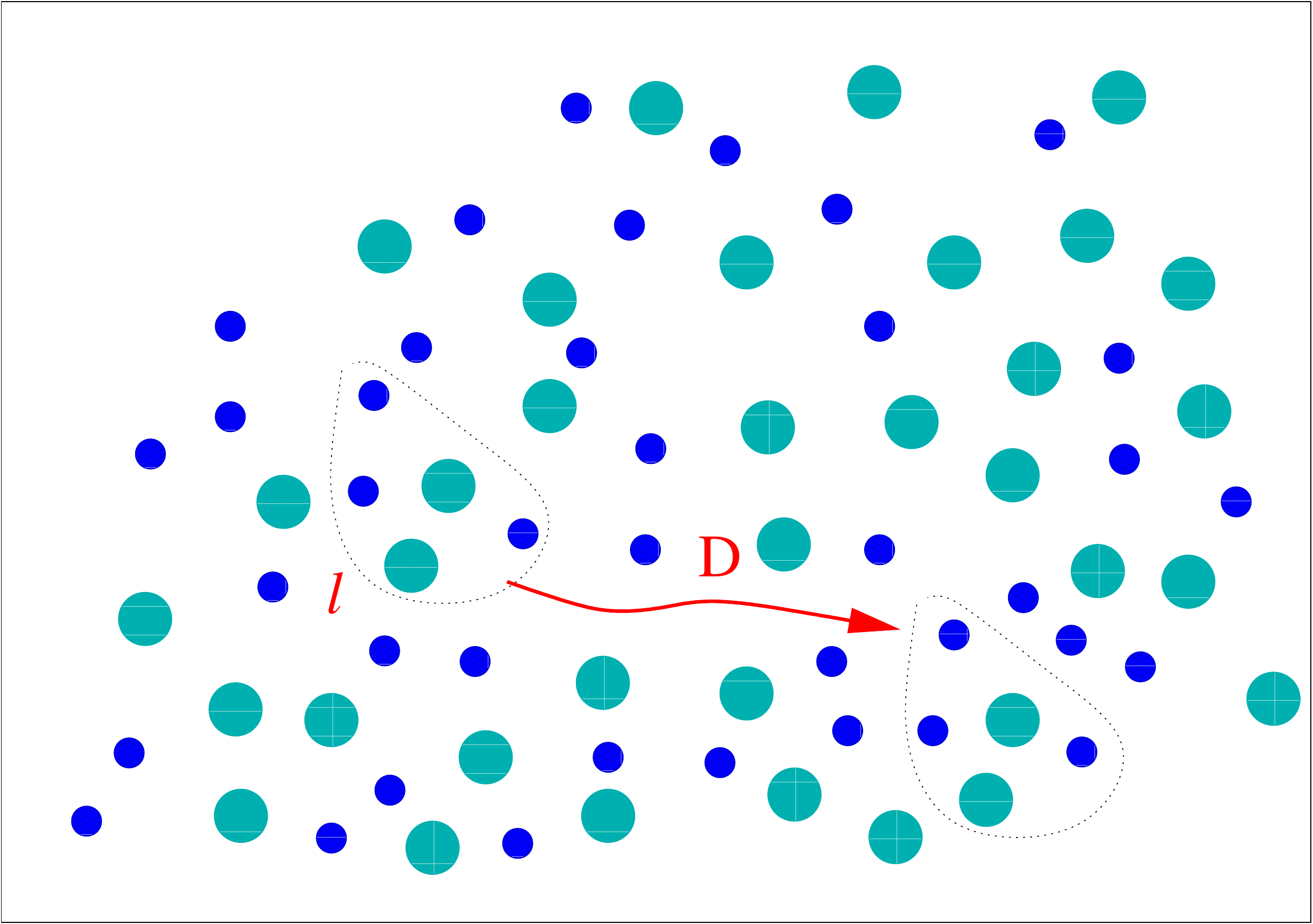}
\end{center}
\caption{ Looking for a  patch that is the same, up to a certain precision, and allowing for rotations. }
\label{avedist1}
\end{figure}

\pagebreak

\section{Scrambled sequences and imperfect tilings}


\subsection{Perfect deterministic sequences}

It is important to note that sub-extensive complexity of patches does not imply that the infinite system is either periodic or quasiperiodic, as characterized by a diffraction spectrum consisting of  $ \delta$ function Bragg peaks. To make this clear, we
review  a particularly well understood  class of systems, namely those which possess `recursive symmetry'.  For ease, we shall consider one-dimensional sequences of two symbols (1 and 0) which may be generated by means
of a substitution rule, wherein a symbol is replaced by a specific word (composed
of 0's and 1's) everywhere it occurs, and then iterating this process.  Generalization to higher 
dimensions, for which building blocks are subdivided, is immediate.  We consider three prototypes
which characterize three types of spatial structure:
\begin{equation}
{\bf Rule \;1:} \; \; \; \; \; \; \; \; \; \; \; \; 1 \rightarrow 10  \; \; \; \; \; \; 0 \rightarrow 10
\end{equation}
Starting from (say) 0, we generate sequences of length $2^{n}$ by iterating 
this rule $n$ times; the result is clearly periodic:
$$
\fbox{0101010101010101010101010101}
$$
 The second possibility is illustrated by 
\begin{equation}
{\bf Rule \;2:} \; \; \; \; \; \; \; \; \; \; \; \; 1 \rightarrow 10  \; \; \; \; \; \; 0 \rightarrow 1
\end{equation}
If we start from 0 as before, we generate strings of length $F_{n}$, where
$F_{n}$ is the $n^{th}$ Fibonacci number:
$$
\fbox{1011010110110101101011011010110110}
$$
This rule produces the {\it Fibonacci sequence}, which is quasiperiodic, with a  discrete Fourier transform consisting of a dense set of $ \delta \;$-functions.
Finally, we consider 
{\bf Rule\; 3:}  
\begin{eqnarray*}
    11 &\to&  1110\\
    10 &\to& 1101\\
    01   &\to& 0010\\
    00 &\to&    0001 
\end{eqnarray*}
Iterating this rule $n$ times produces a string of length $2^{n}$:
\begin{equation}
\fbox{
1110110111100010000111011110110111101101111000101110110100011101}
\end{equation}
This sequence, known as the {\it Rudin-Shapiro sequence}, has a    
discrete Fourier transform with {\it flat} modulus~\cite{Bombieri,thue-morse,Godreche_Luck,Allouche}; and
{\bf Rule\; 3':} 
\begin{equation}
1 \rightarrow 10  \; \; \; \; \; \; 0 \rightarrow 01
\end{equation}
which produces:
\begin{equation}
\fbox{
01101001100101101001011001101001}
\end{equation}
This sequence, known as the {\it Thue-Morse sequence}, also has no  $\delta$-functions
in its Fourier transform~\cite{thue-morse,Godreche_Luck}
The Rudin-Shapiro and the Thue-Morse sequences are examples  of 
``Non Pisot sequences''\cite{Bombieri}.

\begin{figure}[htbp]
\begin{center}
\includegraphics[width=6cm]{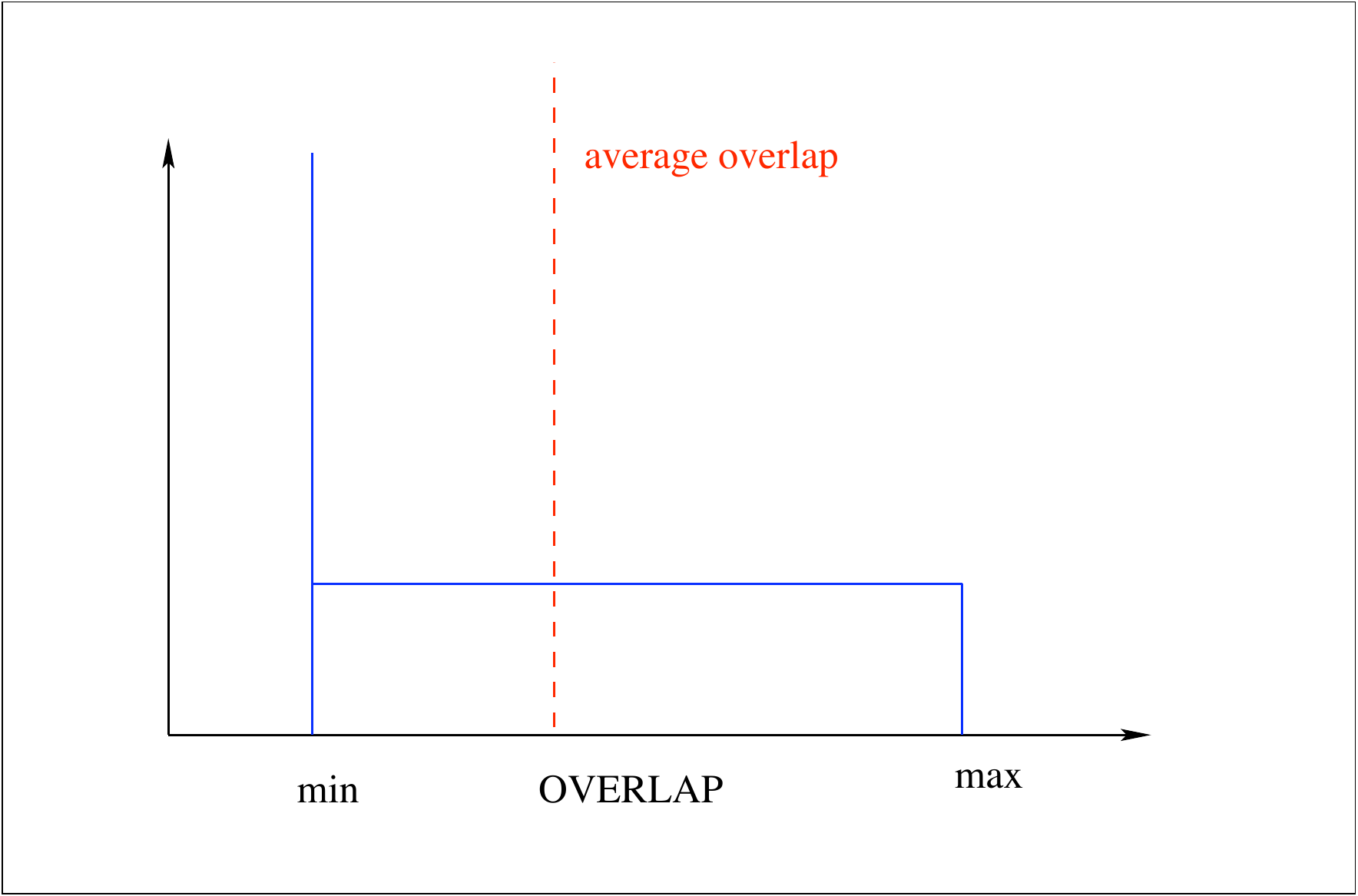}
\end{center}
\vspace{.5cm}
\caption{Probability of overlap of two Fibonacci sequences $P(q)$.}
\label{over}
\end{figure}

Thus, the three examples have different kinds of order:
\begin{enumerate}
\item {\bf Periodic:} Fourier spectrum peaked at the period and its harmonics. Degeneracy $=2$, (entropy $=\ln 2$)  (and, more generally the logarithm  of the  cell size).
\item {\bf Quasiperiodic:} Fourier spectrum is dense set of $ \delta \;$-functions. 
  Two sequences taken at random have an overlap $Q$ distributed according to $P(Q)$ consisting of a $\delta$ function and a uniform tail, as in Fig. \ref{over}:
  the
Parisi function $P(Q)$ is non-trivial. 
The number of subsequences of length $\ell$ (that is, with $\ell$ symbols) appearing in a large sequence is  proportional to $\ell$, (actually $\ell+1$) as is easy to see from the rules above. All patches of size $\ell$ repeat  within a distance of $O(\ell)$. 
The probability that any two subsequences taken at random have a Hamming distance per unit length $< \epsilon$ is proportional to $\epsilon$, and independent of their size $\ell$. 
\item {\bf Non-Pisot sequence:} For the example of the Rudin-Shairo sequence, the modulus of the discrete Fourier transform is {\it flat}, independent of $\omega$. The Parisi function $P(Q)$ is peaked:
two (asymptotically large) subsequences taken at random from the infinite sequence have an overlap that is, with high probability the smallest value possible 
(one-half, in this case). 
Here again, the number of subsequences of length $\ell$ is  also $\propto \ell$, and all patches of size $\ell$ repeat  within a distance of $O(\ell)$.
\end{enumerate}

In going from 1 - 3, some of the conventional (scattering) manifestations of order are lost~\cite{foot11}, despite the fact that the entropies of sequences of types 2 and 3 are still logarithmic, and patches of size $\ell$ repeat much more often than 
$2^\ell$, as they typically would for a random system.   To understand this frequency of repetition, note that a subsequence  of size $\ell$ is completely included within the `descendants' of a single site after $k \propto \log(\ell)$ substitution steps; the  subsequence repeats in the descendant of a neighboring ancestor -
its `cousin $k$-times removed'. {\it This argument is applicable to all systems obtained by substitutions}~\cite{fib,berthe}.

We conclude that  both the crossover  length $\ell_c$  as well as  the complexity length $\ell_1$  diverge in all these deterministic sequences
 -- even though there may be no Bragg peaks in the diffraction spectrum, so that by the conventional measure of scattering such systems would be considered amorphous. 

\begin{figure}[htbp]
\begin{center}
\includegraphics[width=8cm]{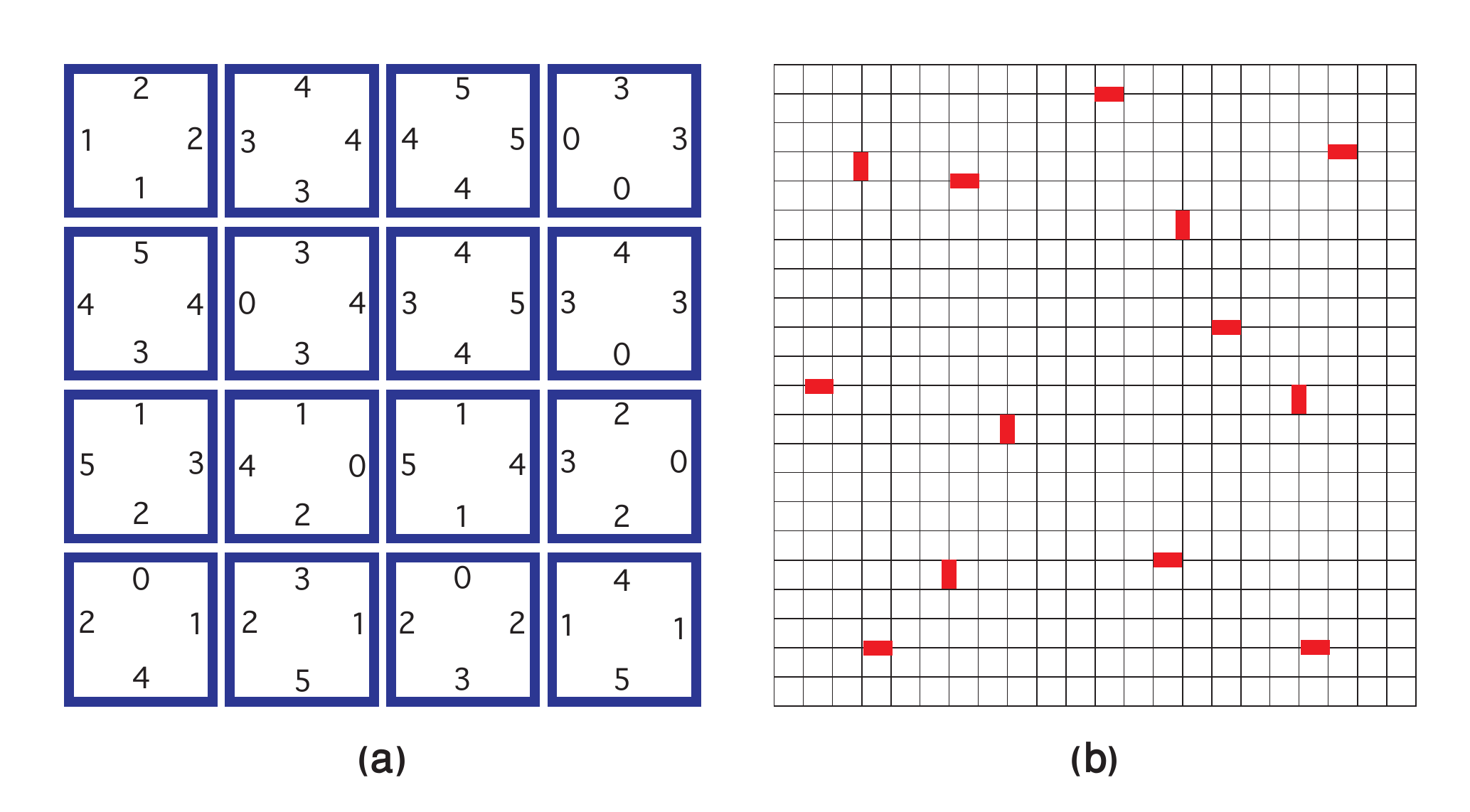}
\end{center}
\vspace{-.5cm}
\caption{(a) A set of Wang tiles with matching rules indicated by numbered edges.  The rule for laying the tiles is that only edges of the same number may abut. (b) States obtained
 from a Monte-Carlo simulation of a Wang tiling quenched from high to low temperature are left with only
`hard' defects  ({\it i.e.} mismatches of adjacent sides); as indicated by the red lines.}
\label{wangtiles}
\end{figure}

These considerations may be generalized to higher dimensions as well.
Perhaps the best known 2D quasiperiodic tiling is the Penrose tiling, but a system with similar properties is a Wang tiling~\cite{Grunbaum}.  In constructing a Wang tiling 
one is required to follow matching rules. One set of tiles with their associated matching rules is shown in Figure \ref{wangtiles}(a): every interface
between tiles has to have the same number on both tiles, a  two-dimensional  domino. 

\begin{figure}[htbp]
\begin{center}
\includegraphics[width=5cm]{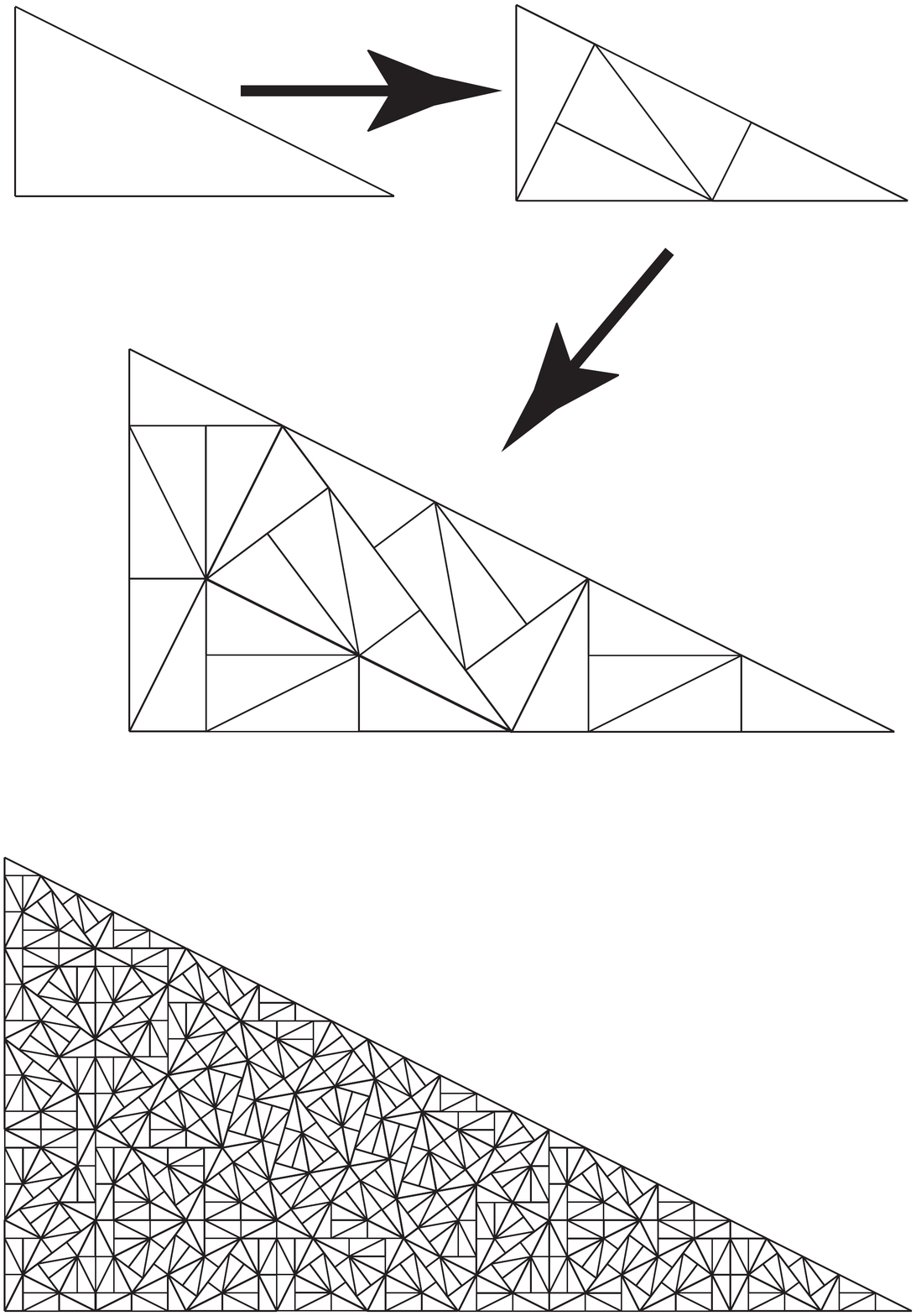}
\end{center}
\caption{The Pinwheel tiling, with infinite correlation length.}
\label{pinw}
\end{figure}

An example of a system  that is statistically isotropic is a tiling of identical triangles~\cite{Radin} 
called the 'Pinwheel' tiling (Fig \ref{pinw}) .
The Pinwheel tiling can be constructed by recursive substitution rules, so patterns of
size $\ell$ repeat (modulo rotation) every ${\cal D} \sim \ell$ in a perfect tiling, as for the 1D sequences 
discussed above.  

Significantly, it is known that any packing which can be generated 
by a recursive substitution rule admits a set of matching rules which only allows
tilings consistent with the recursive symmetry~\cite{Goodman-Srauss}; the
set of matching rules for the Pinwheel tiling is known explicitly~\cite{Radin}.

Two-dimensional tilings of the Non-Pisot type have also been constructed \cite{ClaudeG}. This means that an experimental system having this kind of order would go unnoticed, if the only way to detect quasicrystals order is a diffraction pattern! The strategy we shall outline in the following sections may be used to detect this form of order.

\subsection{{Imperfect sequences:} Finite lengths}

\begin{figure}[htbp]
\begin{center}
\includegraphics[width=6cm]{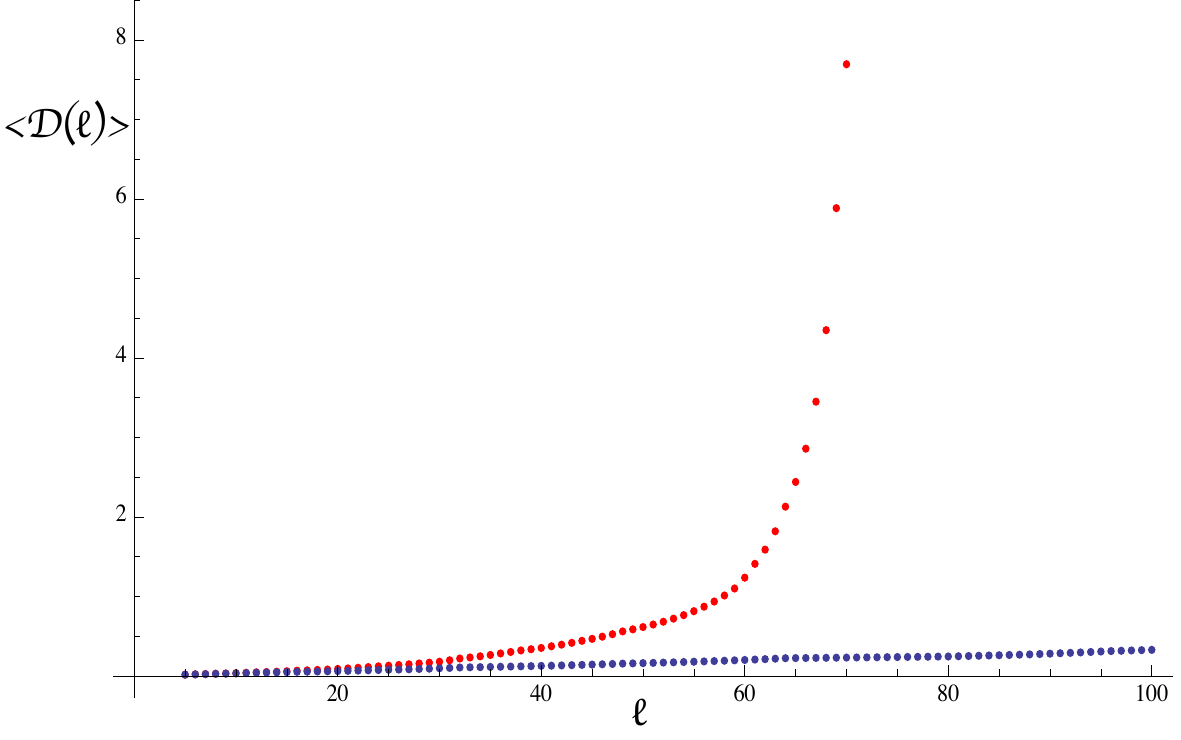}
\end{center}
\caption{ The average distance ${\cal D}(\ell)$ between words of length $\ell$ in   
a perfect Thue-Morse   sequence of length 1000 (blue) and a sequence of length 1000 composed by concatenating
strings of average length 50 randomly drawn from a perfect Thue-Morse sequence
(red).}
\label{avedist}
\end{figure}

The examples of the previous subsection are all systems with infinite correlation lengths. What does an imperfect quasiperiodic system, or 
an imperfect Non-Pisot system look like? The question begins to be tricky, as there is no notion of periodicity in the latter case.
In analogy to a system which has not achieved the ideal glass state, let us consider a sequence constructed by concatenating unrelated subsequences of varying lengths $n$, drawn from a distribution $P(n)$, with average $\overline{n}$. 
For example, the scrambled Thue-Morse sequence  shown below:
\begin{equation}
\fbox{
{\bf   01101 }{\it 0110100101  }{\bf  0011001 }{\it  1001101001}    }
\end{equation}
is the analogue of a polycrystallite, but with micro/Non-Pisot pieces. 
 It is easy to see that strings of length $\ell \ll \overline{n}$ will repeat often, because even in independent subsequences the same string will appear.  
However, if $\ell \gg \overline{n}$, the likelihood of overlap is exponentially small, by the central limit theorem.  This is seen in Figure \ref{avedist},
where the average distance between strings of length $\ell$ is plotted as a function of $\ell$ for a perfect and a randomly concatenated Thue-Morse sequence.  
The two lengths $\ell_c$ and $\ell_1$ are then estimates of the coherence length $\overline{n}$.

Consider next the  example of the  Wang tilings, which  has the virtue that it can 
be expressed as a lattice model of spins~\cite{Leuzzi} (taking $16$ values in the example of 
\cite{Leuzzi}), assigning an energy corresponding to a mismatched edge - that is to
say, the matching rules may be enforced by a suitable Hamiltonian. 
Consequently, such a system can be simulated  at finite temperature (See Figure \ref{wangtiles}(b)) to measure  the growth of the coherence length.
At finite temperature, the Wang tile system has energetic {\it point} 
defects - local violations of the matching rules.
Starting from a high temperature state and annealing down to low temperatures, many easy-to-cure `benign' defects disappear,
leaving the  system in a state with isolated `hard' defects (Figure \ref{wangtiles}(b)) that 
can only be removed by large-scale rearrangements of the order of the interdefect
distance.  Such behavior is also present in the Penrose tilings, where there are benign 
matching rule violations due to local tile flips, and non-trivial configurations of the 
phason field~\cite{coherence} which require large-scale rearrangements to eradicate.  In this case, as we
have verified numerically for the Wang tile system, patches repeat often when much smaller than the distance between 
serious defects, and rarely when they are larger. 
\pagebreak

\section{Particle systems: defining and counting density profiles}

\subsection{Density profiles}

Particle systems are more subtle to compare than number sequences, since they have continuous positions and thermal vibrations.
We will now present a method that  allows us to distinguish and count configurations in principle.  {{Recalling our motivation from glasses, we}} proceed in two steps: first we average out rapid fluctuations to obtain a continuous density profile (a {\it state}),  and then we
count the number of profiles. 

Consider first  a {{snapshot of a}} crystal at finite temperature. It is clear that, strictly speaking, no two patches of the crystal coincide exactly, because of thermal fluctuations in the atomic positions.
Therefore, we consider a density profile averaged over a long (in principle infinite) time:
\begin{equation}
\rho({\bf r}) = \lim_{\tau_\rho \rightarrow \infty} \frac{1}{\tau_\rho} \; \int_{0}^{\tau_\rho}  dt \; \rho({\bf r},t)  \;\;\;\;\; 
\label{average}
\end{equation}
with
\begin{equation}
\rho({\bf r},t) \equiv \sum_a  \delta \left({\bf r} - {\bf r_a(t)}\right) 
\label{average1}
\end{equation}

\begin{figure}[htbp]
\begin{center}
\includegraphics[width=6cm]{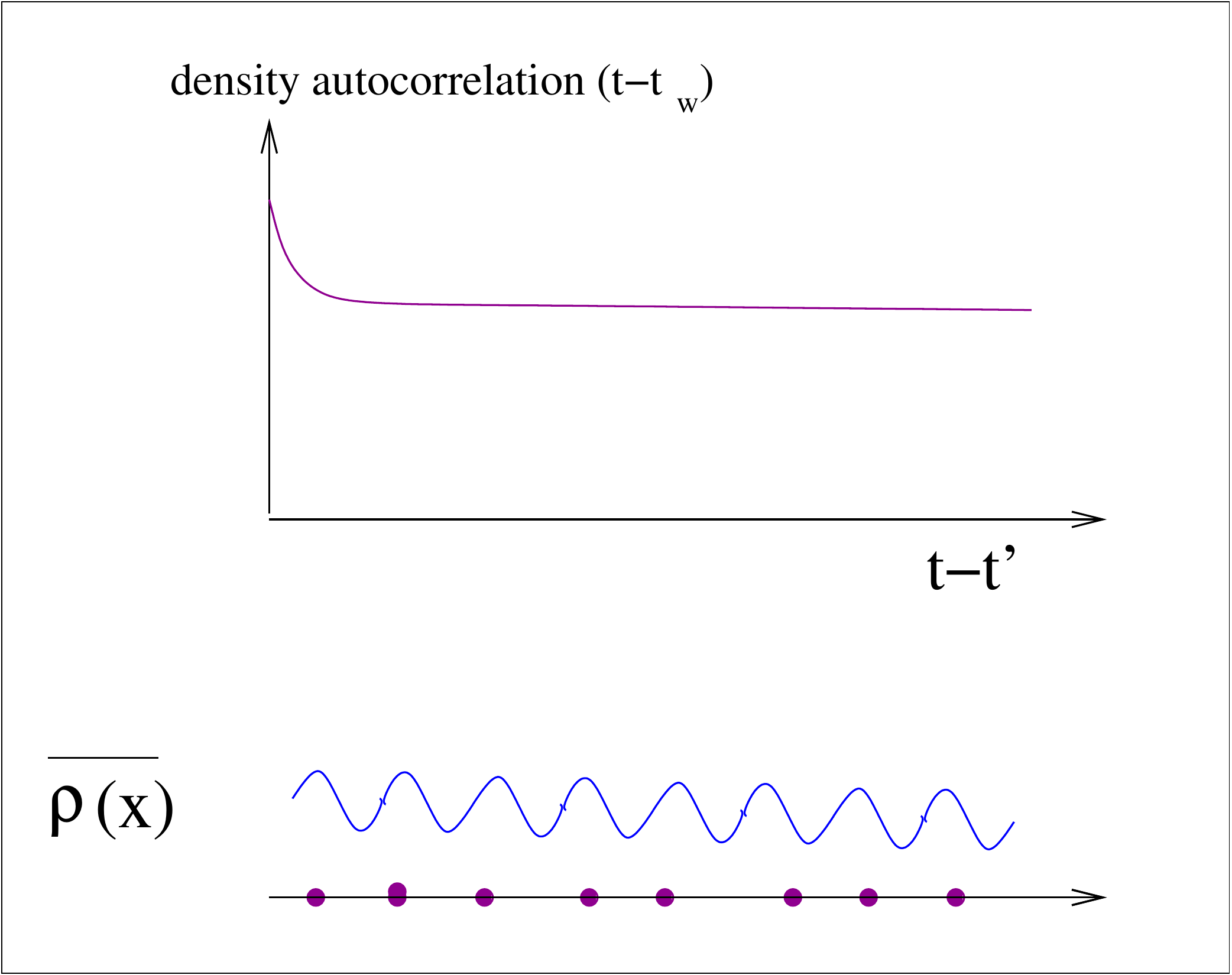}
\end{center}
\caption{ A sketch of the situation in a crystal. At finite temperatures, particles are not in periodic positions. However,
the average density, defined by performing an infinite time average, is indeed periodic.
In this case, we also have the possibility of averaging over spatial repetitions, a procedure automatically performed by the  Fourier transform.
The density-density correlation function has a plateau in time, which lasts forever if there is positional order.}
\label{rho}
\vspace{.5cm}
\end{figure}
In the long-time limit (in this paper, this always means taken {\it after} the thermodynamic limit) , the density tends to a smooth function with a well-defined crystal modulation, as sketched in Figure  \ref{rho}. 
The fact that there are permanent density modulations is reflected in the fact that autocorrelation function
\begin{equation}
C(t,t') = \frac{1}{V}  \int d{\bf r}  \; \left(\rho({\bf r},t) - \bar{\rho} \right)\,\left(\rho({\bf r},t') - \bar{\rho} \right)
\label{correlation}
\end{equation}
has a plateau $q_{EA}$ as $t-t' \rightarrow \infty$:
\begin{equation}
\frac{1}{V} \int d{\bf r} \,\left(\rho(r) - \bar{\rho}\right)^2 = \frac{1}{\tau_\rho^2} \; \int dt \; dt' \; C(t,t') \sim q_{EA} >0
\end{equation}
where {{$$\bar{\rho}=\frac{1}{V}\int d{\bf r}  \; \rho({\bf r})$$}} and $\sim$ denotes in the large $\tau_\rho $ limit.
This defines the Edwards-Anderson parameter $q_{EA}$.

\begin{figure}[htbp]
\begin{center}
\includegraphics[width=6cm]{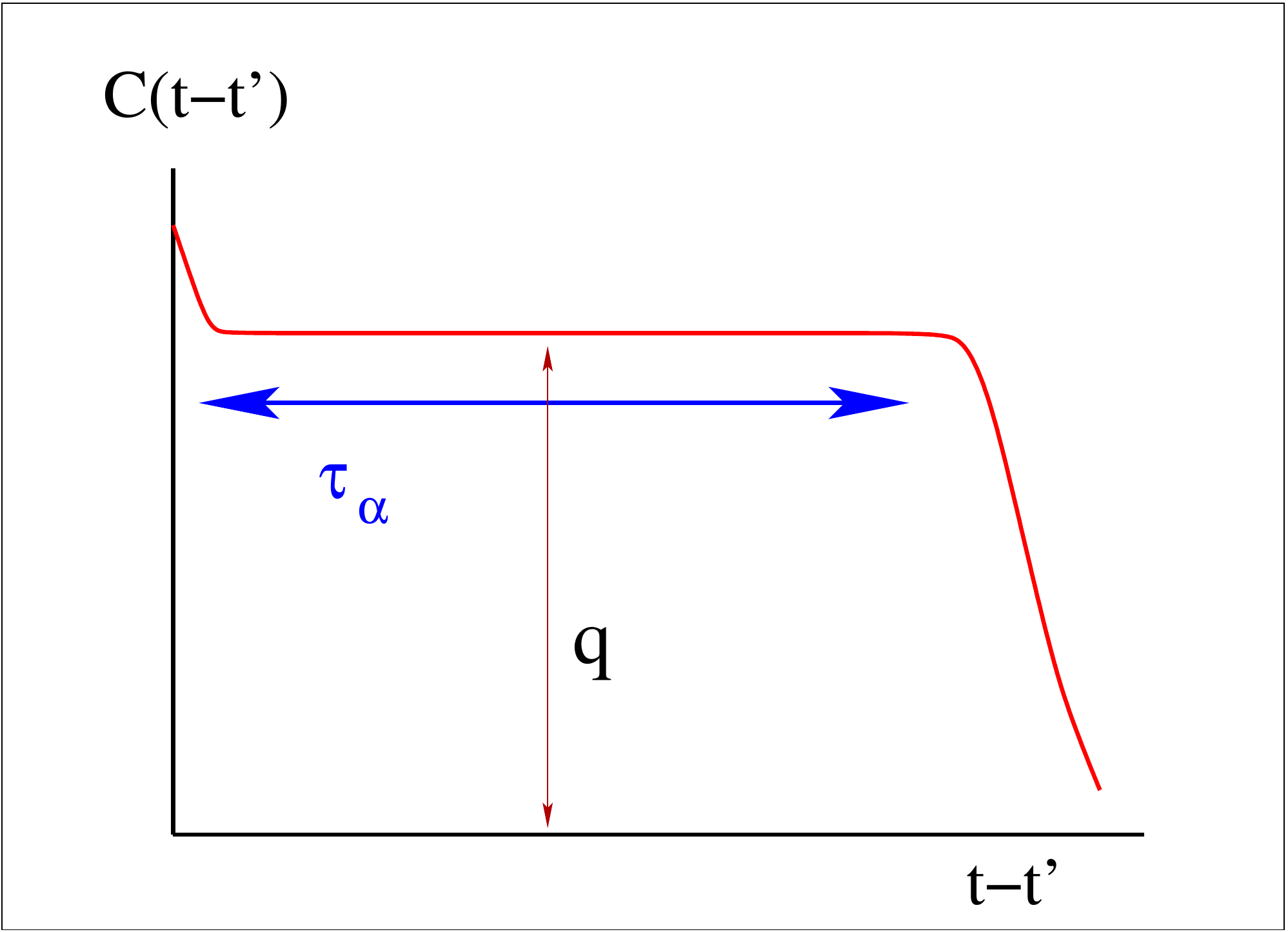}
\end{center}
\caption{A sketch of the autocorrelation function with a two-step relaxation, plotted in logarithmic scale for the times. There is a rapid decrease into the a plateau, whose value is by definition the Edwards-Anderson parameter, followed by a fall to zero at a longer time scale $\tau_\alpha$.  The slow timescale $\tau_\alpha$ for an equilibrium system depends on the temperature,
and grows as the glassy phase is approached. Out of equilibrium, $\tau_\alpha$ grows  with the `waiting' time after preparation, and saturates if and when
equilibrium is reached.    }
\label{plateau}
\vspace{.5cm}
\end{figure}

An ideal amorphous solid should have permanent, non-periodic  density modulations. By this we mean that, if we define a density
profile as in equation  (\ref{average}), we obtain a function that may be  neither constant nor periodic nor quasiperiodic.
Such solids might or might not strictly speaking exist.  What we know in practice  is that a supercooled liquid has a relaxation time $\tau_\alpha$,
much larger than the microscopic time $\tau_m$, 
during which an amorphous  density profile remains stable,  so that the autocorrelation function (\ref{correlation}) has a plateau as in Figure \ref{plateau}. Only
if $\tau_\alpha \rightarrow \infty$ do we have a true amorphous solid. 

For finite $\tau_\alpha$, we may define
 a density profile for a supercooled liquid configuration as in  equation (\ref{average}), but with the average taken over a finite time $\tau_\rho$ such that $\tau_m \ll \tau_\rho \ll \tau_\alpha$.
This brings about an essential limitation on the precision with which both $\rho(\bf r)$ and $\bf r$ are defined. 
The density resolution  is of the order of, but  inevitably not better than:   
\begin{equation}
\epsilon (\tau_\rho) \sim \left[ \frac{1}{V \tau_\rho ^2} \int d{\bf r} \; dt (\rho^2 ({\bf r},t) - \rho^2({\bf r}) )\right]^\frac{1}{2} = \left[ \frac{1}{\tau_\rho^2} \int_0^{\tau_\rho}
dt \; dt' \; C(t,t') - q_{EA} \right]^{\frac{1}{2} }
\end{equation}
The precise manner in which the resolution $\epsilon(\tau_\rho)$ goes to zero
depends on the way that $[C(t-t')-q_{EA}]$ goes to zero on approaching the plateau: if it does so in an integrable form, then $\epsilon(\tau_\rho) \sim (\tau_\rho)^{-\frac12}$. 

The resolution of ${\bf r}$ is in turn limited by the fact that the density is generated by time-dependent $\delta$ functions , corresponding to particle positions.
Suppose we divide the total volume ${\cal V}$ in cubes of side $ \Delta {\bf r}$. A particle travelling with velocity ${\bf v}$ will visit ${\bf v} \tau_\rho /  \Delta {\bf r}$ cubes
in time $\tau_\rho$. Estimating ${\bf v} \sim \sqrt{k_B T}$, and considering that there are $N$ particles, a total of  order $ N \sqrt{k_B T} \tau_\rho /  \Delta {\bf r}$
are visited. If we want this to be larger  than  the total number ${\cal V}/    [\Delta {\bf r}]^D$ of cubes, we get the condition:
\begin{equation}
 \Delta {\bf r} (\tau_\rho) \sim \left[ \sqrt{k_B T}\tau_\rho \rho\right]^{\frac{1}{D-1}}
\label{smo}
\end{equation}
For large $\tau_{\rho}$, $ \Delta {\bf r} (\tau_\rho)$ is exceedingly small, and we shall assume throughout that we have discretized space or smoothed the average density functions
up to a distance of this order.  In what follows we shall, in addition, sample $\rho$ at even larger spatial intervals, but this we shall do after having 
discretized space  at the scale (\ref{smo}).

In realistic  situations, the above time-averaging procedure may be impractical.  Two alternatives to this procedure are \\ $\bullet \;\;$
Identifying positions  modulo some fixed tolerance that we know {\it a priori} to be appropriate, for example a fraction of the inter-particle spacing.\\
$\bullet \;\;${{Instead of comparing the averaged configurations, we may omit the time-averaging and compare the}} inherent structures of configurations, that is,
the positions reached after a quench to $T=0$. Using inherent structures, however, is not free of artifacts and ambiguities: two inherent structures that differ 
in  buckling modes that flip-flop on a timescale $<\tau_{\alpha}$ should be treated as being the same, while a naive counting will treat them as different.
If the density of such localized flip-flops is finite, one gets an entropy that { may} be absent in a correctly time-averaged procedure. 

\subsection{Correlation lengths}

The preceding procedure yields a density profile that has averaged out -- to the extent possible -- all rapid degrees of freedom, such as 
particle-vacancy pairs, phonons, particle rattlings. Our problem now is how to count the resulting continuous profiles, or, alternatively, how to identify congruent patches.
This situation is already familiar in a different context: it arises when we wish to determine  the entropy of trajectories of a dynamical system, in which setting it has been 
well discussed both in principle and in practice. 
In the case of  dynamical systems, we first discretize phase-space in cubes of size $\epsilon$ and time in intervals $\tau$. We consider pieces of trajectories $d$ time-intervals
long, and define the Kolmogorov entropy {{on the basis of the probability $P_\epsilon(i_1,...,i_d)$ that at time $\tau$ the trajectory lies in the cube $i_1$, at time $2\tau$ it lies in the cube $i_2$, ..., and at time $d\tau$ it lies in cube $i_{d}$}}, as \cite{Paladin}:
\begin{equation}
K_1= - \lim_{\tau \rightarrow 0} \lim_{\epsilon \rightarrow 0} \lim_{d \rightarrow \infty} \frac{1}{\tau d }  \sum_{i_1,...,i_d} P_\epsilon(i_1,...,i_d)
\ln P_\epsilon(i_1,...,i_d)
\label{KS}
\end{equation}
More generally, we may define all the Renyi complexities:
\begin{equation}
K_q= - \lim_{\tau \rightarrow 0} \lim_{\epsilon \rightarrow 0} \lim_{d \rightarrow \infty} \frac{1}{\tau d (q-1)}  \ln \left(\sum_{i_1,...,i_d} P_\epsilon(i_1,...,i_d)^q \right)
\label{Renyi}
\end{equation}
Note that $d$ goes to infinity {\it before} $\epsilon$ and $\tau$ go to zero, a fact that shall turn out to be important.

\begin{figure}[htbp]
\begin{center}
\includegraphics[width=6cm]{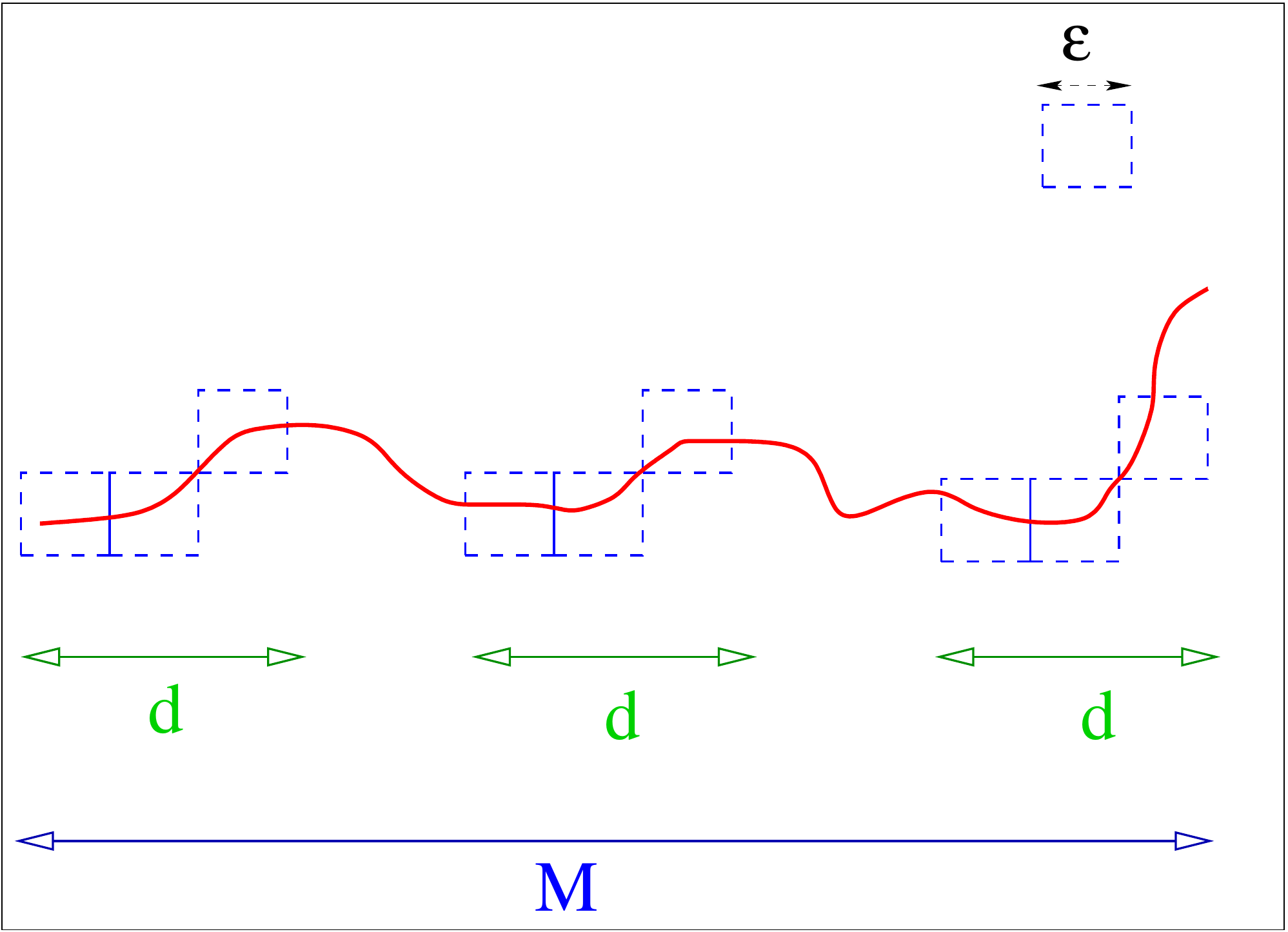}
\end{center}
\caption{ Discretization of trajectories. A sequence is repeated three times at this level of precision, but each one will be counted 
as distinct at a smaller $\epsilon$.}
\label{Grass}
\vspace{.5cm}
\end{figure}

\begin{figure}[htbp]
\begin{center}
\includegraphics[width=8cm]{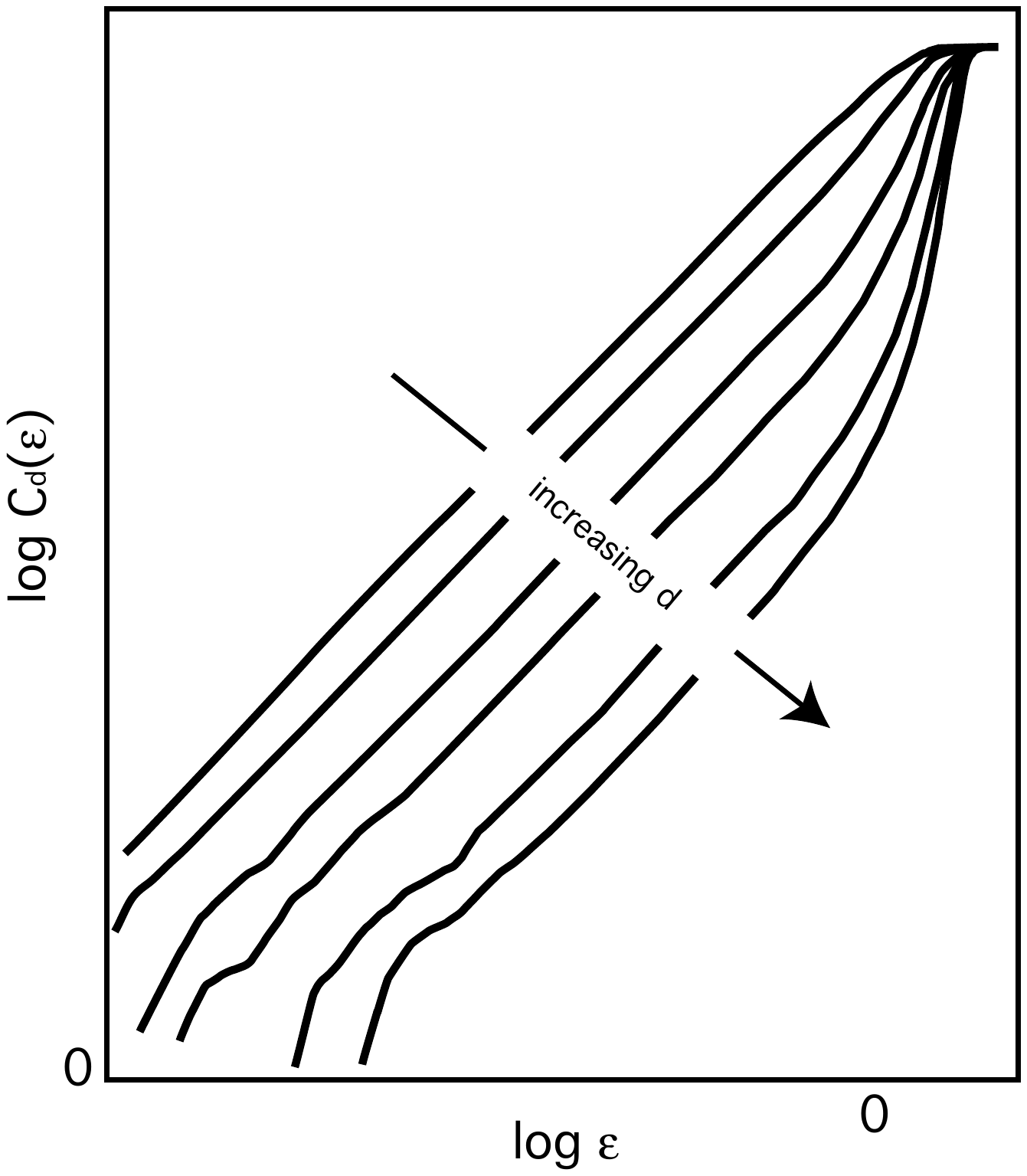}
\end{center}
\vspace{-3cm}
\caption{ A sketch of the plots of the logarithm of patch frequency versus  precision $\epsilon$, for different lengths $d$ (see Ref \cite{Grassberger}).
The Grassberger-Procaccia procedure takes advantage of the scaling with $\epsilon$ to extrapolate results
from relatively large $\epsilon$, to  small $\epsilon $, a limit in which no coincidences would in practice be found. }
\label{GP}
\end{figure}

In practice, one may  proceed as proposed by  Grassberger and Procaccia~\cite{Grassberger}: we consider a very long sequence of $M$ times, and for every patch of length $d$
labeled by its starting point $i$ (Fig. \ref{Grass}) , we count the number of patches $n_i^{(d)}(\epsilon)$ that coincide with it -- up to precision $\epsilon$.
The entropies are estimated by 
\begin{equation}
K_q \sim - \lim_{\tau \rightarrow 0} \lim_{\epsilon \rightarrow 0} \lim_{d \rightarrow \infty} \frac{1}{\tau d (q-1)  } \ln \left\{ \frac{1}{M} \sum_i [n_i^{(d)}]^{(q-1)} \right\}
\label{GP1}
\end{equation}
and in particular:
\begin{equation}
K_1 \sim - \lim_{\tau \rightarrow 0} \lim_{\epsilon \rightarrow 0} \lim_{d \rightarrow \infty} \frac{1}{\tau d  } \left\{ \frac{1}{M} \sum_i \ln [n_i^{(d)}] \right\}
\label{GP2}
\end{equation}
This way of putting the problem is practical, because we may extrapolate the results from finite $\epsilon$ (for which we have many coincidences) to smaller
$\epsilon$. This is illustrated in Fig. \ref{GP} (a sketch). 

The fact that an entropy may be defined and computed  may be understood if we assume that \cite{Paladin} the dependence on $\epsilon$ (for small $\epsilon$) 
is subdominant in $d$:
\begin{equation}
\frac{1}{M} \sum_i [n_i^{(d)}]^{q} \sim e^{\tau d K_{q+1}} \epsilon^{\phi(q)} = e^{\tau d K_{q+1} + \phi(q) \;  \ln \epsilon } 
\label{subdo}
\end{equation}
for some $\phi(q)$. 
 As we shall discuss later, this means that  $d$ goes to infinity keeping $|\ln \epsilon|/ d$ small.
\begin{figure}[htbp]
\begin{center}
\includegraphics[width=6cm]{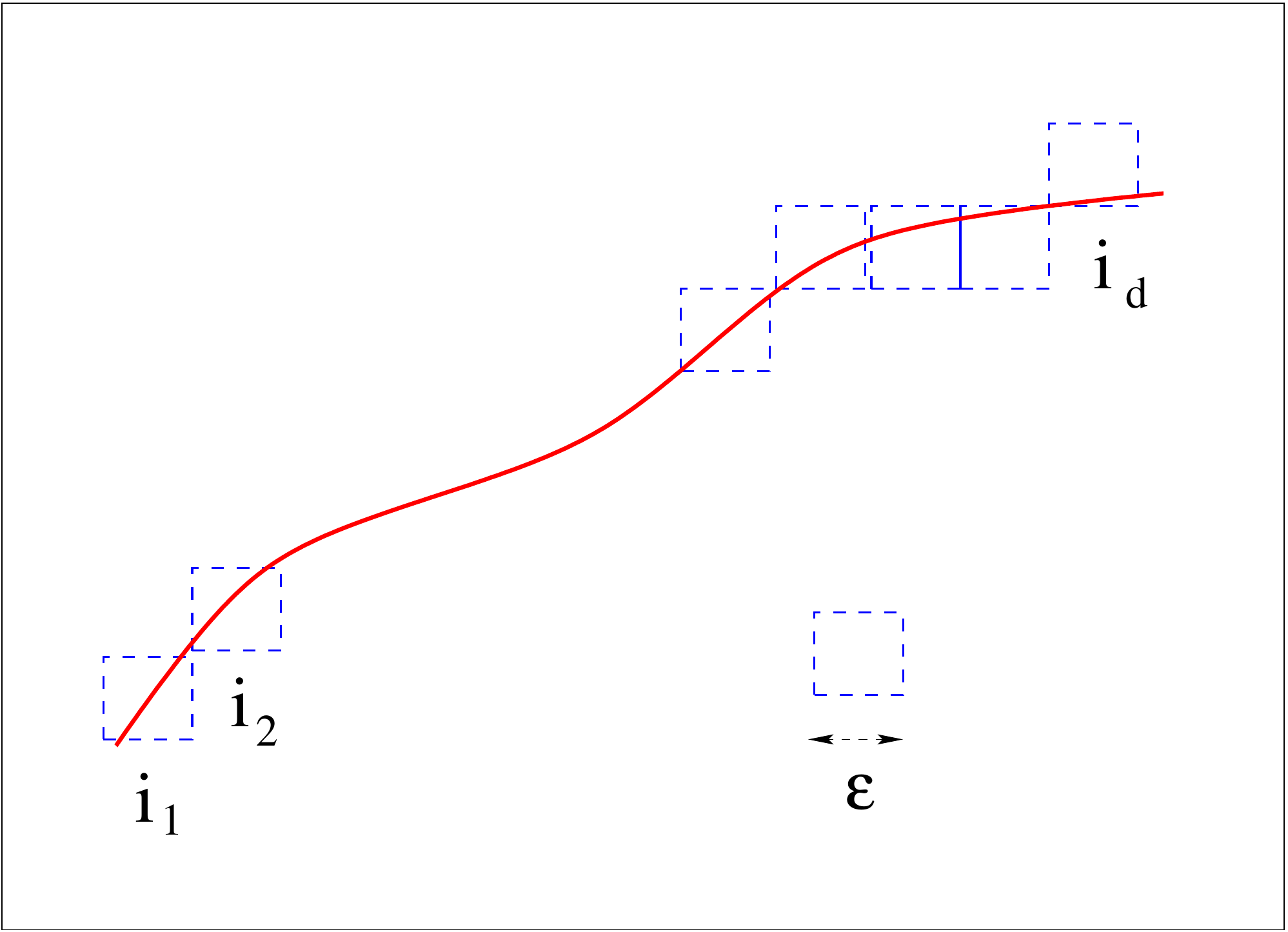}
\end{center}
\caption{ Counting trajectories in a discretized space and time.}
\label{KSi}
\vspace{.5cm}
\end{figure}

\begin{figure}[htbp]
\begin{center}
\includegraphics[width=6cm]{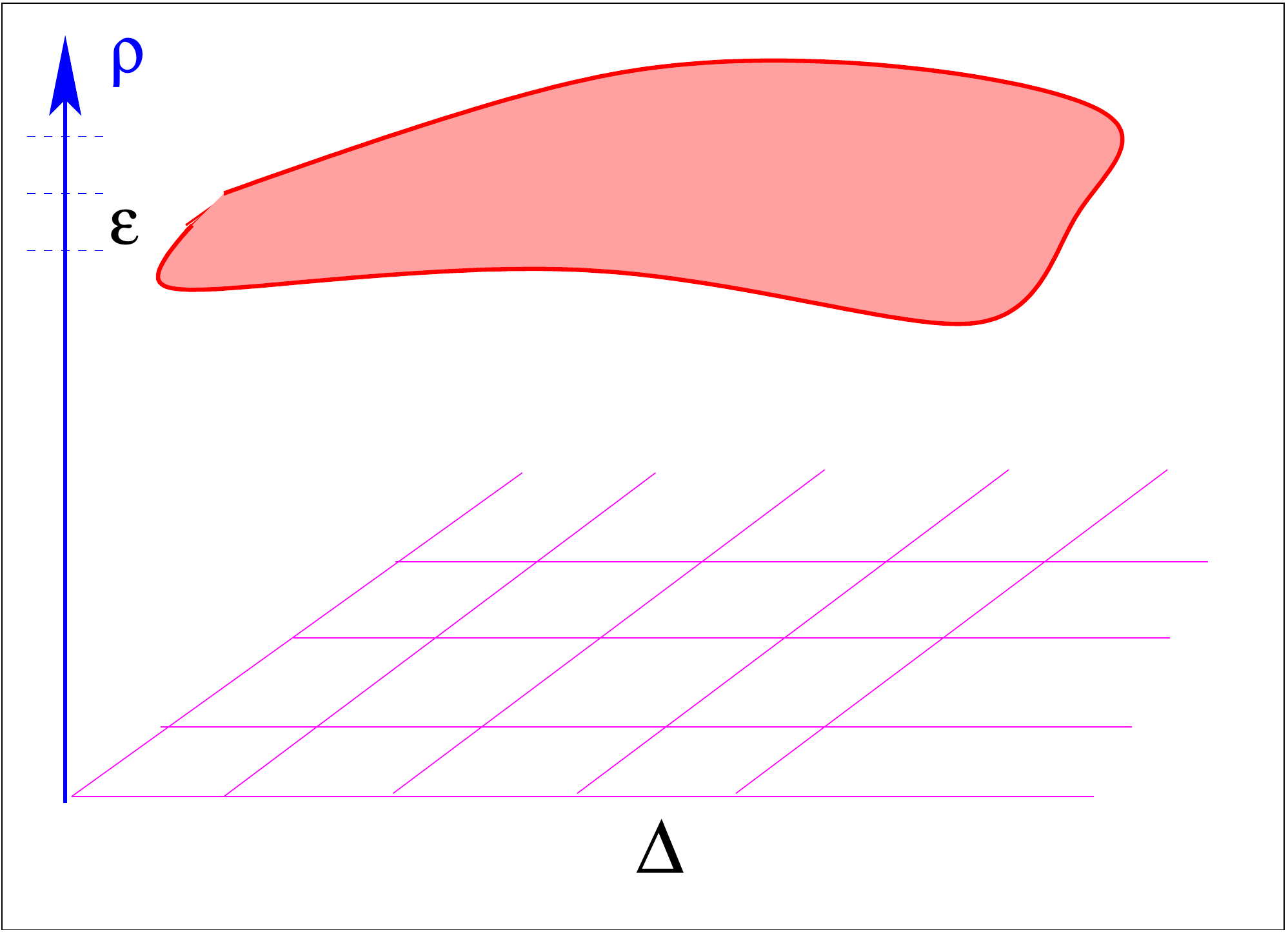}  
\end{center}
\caption{ The analogue of a trajectory is the density profile. The density plays the role of the position in a dynamical system, and space the role of time.}
\label{dens}
\vspace{.5cm}
\end{figure}

We may easily adapt this procedure to count density profiles. We consider average profiles $\rho({\bf r})$ in $D$-dimensional space (Fig. \ref{dens}). We discretize space in some manner,
with a typical grid spacing $ \Delta$ playing the role of the time-interval $\tau$ in a dynamic system~\cite{foot22}, and we also discretize  $\rho$ in intervals of size $\epsilon$.
Here again, we may proceed as follows. 
We consider a large sample containing $M$ different patches ${\bf a_1, ..., a_M}$ each of volume $V$. 
We look for the number of coincidences between a patch and all others,  centered about any site and allowing for rotations. Two patches coincide
if the { sum} of all the differences in the value of $\rho$ between corresponding sites is smaller than $\epsilon$.
 Denoting by $n_a^{(V)}(\epsilon) $ the number of patches that coincide in this way with the patch ${\bf a}$, we have:
 \begin{equation}
K_q \sim - \lim_{ \Delta \; \rightarrow 0} \lim_{\epsilon \rightarrow 0} \lim_{V \rightarrow \infty} \frac{1}{ (q-1) V    } \ln \left\{ \frac{1}{M} \sum_a [n_a^{(V)}(\epsilon)]^{(q-1)} \right\}
\label{GP3}
\end{equation}
and in particular:
\begin{equation}
K_1 \sim - \lim_{ \Delta \; \rightarrow 0} \lim_{\epsilon \rightarrow 0} \lim_{V \rightarrow \infty} \frac{1}{   V     } \left\{ \frac{1}{M} \sum_i \ln [n_a^{(V)}(\epsilon)] \right\}
\label{GP4}
\end{equation}
Here and in what follows we shall for (conceptual, though not practical)  simplicity keep the limits $ \Delta \; \rightarrow 0$ and
$\epsilon \rightarrow 0$, although they are probably unnecessary provided that one takes $V \rightarrow \infty$.

 The Renyi  complexities are at most extensive  ($K_q = O(1)$ as $V \rightarrow \infty$), or subextensive, and  then   $K_q \rightarrow 0$.  Writing equation (\ref{GP3}) in the form
 \begin{equation}
 e^{-q V   \; K_{q+1}} \sim  \frac{1}{M}  \sum_a [n_a^{(V)}(\epsilon)]^q   =  \frac{1}{M} \sum_a e^{q V  \, [\frac{\ln n_a} {V  \;}  ] }    
 \label{coco}
\end{equation}
(where the symbol `$\sim$' means ` in the limit $  { \Delta \; \rightarrow 0} \; {\epsilon \rightarrow 0} \; {V \rightarrow \infty} $')
suggests that a large  deviation function be defined for the variable $\beta Y  = -\frac{\ln n_a} {V } $ (the sign and the factor
$\beta$ entering the definition of $Y$ are purely conventional).  Introducing the large deviation function $G$ through:
\begin{eqnarray}
{\cal P} \left(\beta Y  = -\frac{\ln n_a} {V \;}  \right) &\sim&  e^{V  \; G(Y)} = \frac{1}{M} \sum_a   \delta \;(V  \; \beta Y  + \ln n_a) \nonumber \\
&=& \int dq \,  \frac{1}{M} \left[ \sum_a e^{q  V   \; \left\{ (\frac{\ln n_a} {V  \,}  ) + \beta Y  \right\} }    \right] \nonumber \\
&=& \int dq \; e^{ q V    \, [\beta Y  -K_{q+1}]}
\label{oop}
\end{eqnarray}
Thus, the Renyi complexities may be thought of as (a form of) a Legendre transform of  the large-deviation function $G(Y)$. We may evaluate the integral in (\ref{oop}) by saddle-point
\begin{equation}
G(Y)= q^* [\beta Y  - K_{q^*+1}] \;\;\;\;\; with \;\;\; \;\;\beta Y = \left.\frac{d }{dq}  \left[q K_{q+1}\right]\right|_{q^*} 
\label{oop1}
\end{equation}

Before proceeding, let us be more precise about the dependence on $\epsilon$. If we assume, as in (\ref{subdo}),  that
(\ref{coco}) may be written in the limit of small but non zero $(\Delta, \,\epsilon)$ and large but not infinite V as:
\begin{equation}
  \frac{1}{M}  \sum_a [n_a^{(V)}(\epsilon)]^q   \sim  e^{-q V  K_{q+1} + ... } \;\; \epsilon^{\phi(q)}
 \label{coco1}
\end{equation}
for small $\epsilon$. Note that there might be a correction intermediate in order $V$ between the leading one and the one dependent on $\epsilon$. 
 If we  work under the condition
\begin{equation}
\frac{\phi |\ln \epsilon| }{V K_{q+1} } \ll 1  \;\;\; \rightarrow \;\;\;  \epsilon >>  e^{-V K_{q+1}/\phi}
\label{condi}
\end{equation}
 we may refine equation (\ref{oop1}) to read:
\begin{equation}
\ln {\cal{P}}(Y) =   V  \Delta \; \; G(Y) + \ln\epsilon \; G_o(Y)  \;\; \;\;\;\; with  \;\;\; \;\;\;    G_o = \phi(q^*(Y)) 
\label{asym}
\end{equation}
with $q^*(Y)  $ given in $(\ref{oop1})$.
From this equation we read the approach, as $\epsilon \;  e^{K_{q+1}  V} >>0$, of the large-deviation function $G$ to its large-$V$ limit.
Let us stress that we do not know what exactly the dependence of entropy on $\epsilon$ might be like in a  true glassy system, and this is one of
the interesting questions an experimental/numerical investigation should answer. 

\subsection{ Direct measurement of the complexity}

The  (patch, or block) complexity 
$\Sigma^V $, is defined as the log of the number of different kinds of states for a given patch size V, per unit volume.  (Here, and in what follows, $\Sigma$
without supraindex  stands for the large $V$ limit).
 Because each state $\alpha$  is repeated $n_\alpha$ times, we have to divide by $n_\alpha$
 to count each type once

\begin{equation}
 \Sigma(Y)  =  \lim_{ \Delta \; \rightarrow 0} \lim_{\epsilon \rightarrow 0} \lim_{V \rightarrow \infty} \frac{1} {V  }\ln \left\{ \frac{{\cal P}[n(\epsilon)]}{n(\epsilon)} \right\} =  \beta Y  + G(Y) \;\;\;\;\;\; and  \;\;\;\;\; \beta Y  \equiv- \frac{\ln n}{V  } 
 \label{complexity}
 \end{equation}
 From equations (\ref{complexity}) and (\ref{oop}) we obtain the  useful  expression:
 \begin{equation}
 e^{-(q-1) V K_q} = \int \; dY \; e^{\Sigma(Y)- \beta q Y}
 \label{iii}
 \end{equation}
 
 In practice, one would work at different values of $ \Delta \;$ and $V$, and make histograms of the probability that patterns repeat with frequency $n(\epsilon)$ at given 
 precision $\epsilon$. Plotting in terms of the normalized logarithm of this frequency  $\beta Y =\ln\left(\frac{n(\epsilon)}{V  }\right) $
one should obtain a set of curves that scale correctly as $\Sigma (\epsilon, V )  = \Sigma(Y) + \frac{\ln\epsilon}{V }  \; G_o(Y)$. 

  The content of  the functions  $G(Y)$, $\Sigma(Y)$ and  $K_q$'s is the same.  They encode  important information  to distinguish different scenarios for the amorphous state.

\section{Long range order}

\subsection{Amorphous order  vs. Weak-periodicity}

The property  of weak periodicity,   introduced by Aubry, is defined as follows~\cite{Aubry}: for patches
of size $V$, there is a distance $R(V,\epsilon)$ along the sample such that in an infinite sample every patch will recur -- up to precision $\epsilon$ -- at a distance  closer than $R(V,\epsilon)$.  Note that nothing is said about the dependence of $R(V,\epsilon)$ with $V$. What this definition excludes 
is the possibility that   for any given type of patch ${\bf a}$, there might  be, in an infinite sample,  arbitrarily
 large  ${\bf a}$-free regions, something that would clearly happen in a random sample.
 Aubry~\cite{Aubry} showed that, under reasonable assumptions, the ground-state configuration of a particle system with short range interactions is weakly periodic.
 
 In this paper, we are defining a  length $\ell_1$ for amorphous order also based on patch repetition which  is,   although conceptually related,   not equivalent. 
 Central to our argument is the necessity of time-averaging, which allows us to compare average-density profiles. We are measuring {\it complexities}  -- based on the multiplicities of these averaged profiles -- rather than {\it entropies},  which are related to the multiplicities of configurations. Time averaging  allows us to work at non-zero temperature, a situation in which weak periodicity of configurations necessarily breaks down.  
 
 The definition we follow in this paper of perfect order, implying an infinite correlation length, is a 
   situation in which {\it almost every patch recurs often}, the recurrence distance being  less than exponential in the patch's  size. This is related to the vanishing of the complexity, and it does {\it not} exclude the possibility that there will be arbitrarily large regions where some  patch
  is absent. 
  
  Is it possible that we could use the notion of weak periodicity applied, rather than to configurations,  to time-averaged density profiles?
 The answer is that, even in that case, the two conditions (vanishing complexity and weak periodicity) are distinct.
To understand this,   consider first the case of a quasicrystal at low but non-zero temperature, for example the Wang-tile system discussed above.
   In equilibrium, defects will appear and disappear  continuously, just as in a crystal. These `benign' defects will be averaged out in our 
   procedure, so we may safely ignore them. However, there are also `serious' defects having low energies that cannot be eliminated
  with a finite number of rearrangements (the best known being the `decapods' of the Penrose tilings).  These are long-lived defects, and will therefore show up in a time-averaged 
  density function. Moreover, they have, in equilibrium, finite probability if their energy is finite; as is the case in the  Wang-tile and Penrose system above.
  
  Denote by $f_\alpha$ the free energy density  of the metastable state $\alpha$ (measured with respect to the one $f_o$ of a perfectly ordered system)
   defined as having a  fixed set  of `serious' defects,  and including all rapid thermal fluctuations. $f_\alpha$ is computed from the partial Gibbs-Boltzmann distribution, keeping `serious' defects fixed.
    Denote  $ {\cal N} = e^{{\cal{V}} \Sigma(f)}$
 the number of different such metastable states of  free energy density $f$ for a sample of total volume ${\cal{V}}$.
  We may write the partition function at low temperature as:
  \begin{equation}
  Z = e^{-\beta {{\cal{V}}} f_o} \;   \int df \; e^{{\cal{V}}(\Sigma(f) - \beta f)}
  \label{froze}
  \end{equation}
  The integral may be evaluated for large ${\cal{V}}$. If $\left. \frac{d\Sigma}{df}\right|_{f=0} > \frac{1}{T}$ a saddle point with $f>0$ dominates, 
  there is a finite density of `serious' defects in equilibrium. (As we shall see below,  a finite density of defects brings the system back
  to the liquid phase, in this case  because defects are  so close that they are
 may be mutually annihilated with finite rearrangements, and  have no longer infinite life). If, on the other  hand $\left. \frac{d\Sigma}{df}\right|_{f=0} < \frac1T$, the system is frozen in $f\sim 0$, order is not destroyed
  because the density of defects tends to zero, although the total number present may still be large.
   In the latter case we have an ordered system with infinite complexity-related correlation length, but {\it no} weak periodicity (even for a time-averaged configuration)   since the `serious' defects spoil it. 
  
  Although we have chosen to discuss this situation in terms of the Wang tile system, it is also relevant for the (more hypothetical) `ideal glass' scenario at finite  temperature, as we shall see.

\vspace{.2cm}

\subsection{Infinite lifetime and diverging lengths.}

\vspace{.2cm}

We shall now argue that if  a system of particles with short-range smooth interactions has a  state with permanent density modulations, then these density profiles necessarily have subextensive complexity, and hence there is, by our definition,   long-range amorphous order. This is also true for the zero-temperature limit of a system with super-Arrhenius timescale behavior.
Our reasoning is not rigorous, but we believe it may be made so.  The argument is based on an elementary observation: {\it it is not possible to have exponentially many  metastable states  
all of which have  a divergent  lifetime}.

\begin{figure}[h]
\begin{center}
\includegraphics[width=8cm]{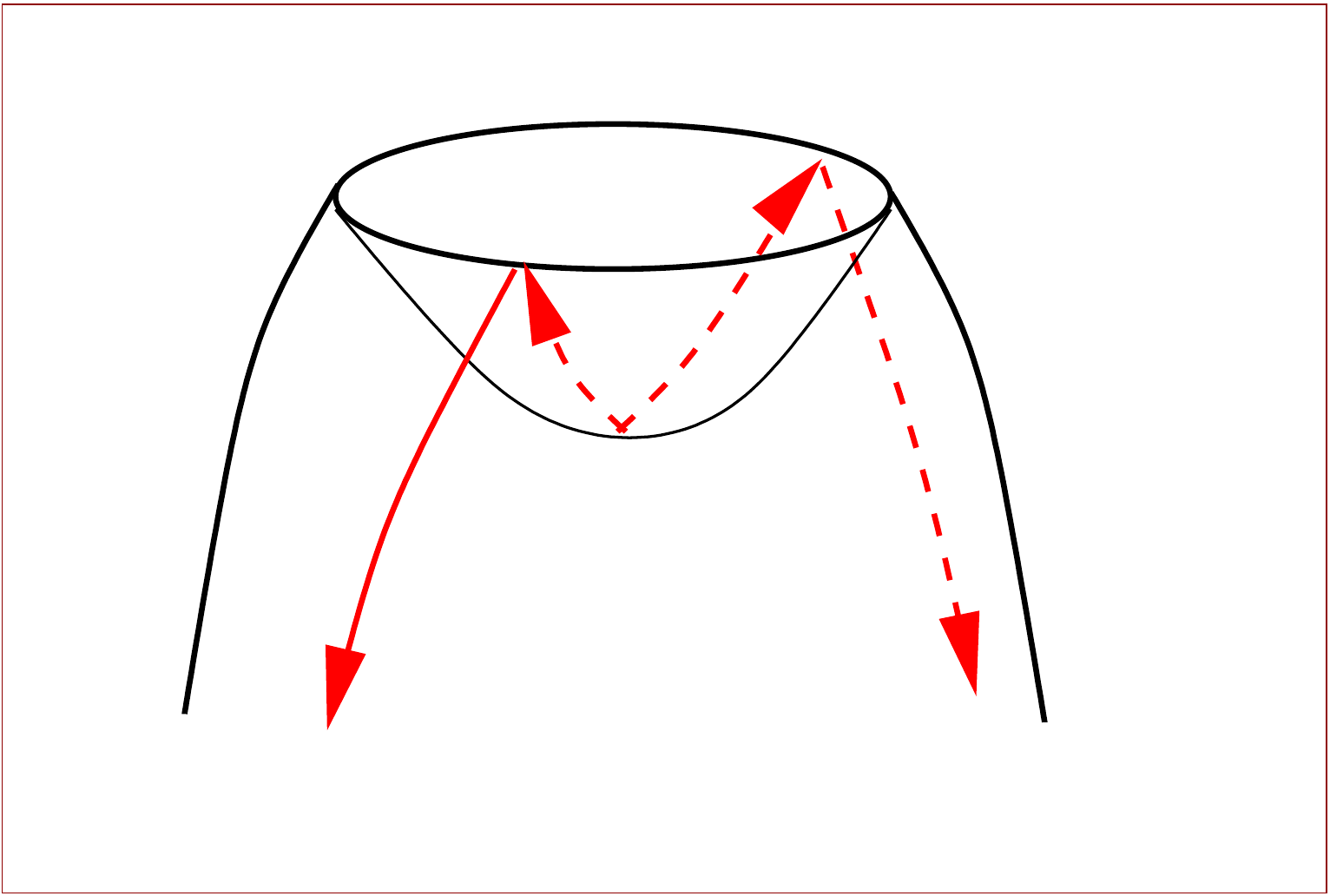}
\end{center}
\caption{Two trajectories escaping a crater.}\label{mexican}
\end{figure}

A situation that arises in supercooled liquids is that a system has many options of phases for nucleating. 
The question then  is: does this multiplicity  increase the probability of nucleation into  a different phase?   In other words, is there an `entropic drive' for nucleation away from a state, 
leading the  system to another state which, though  individually not necessarily more favorable,  belongs to an exponential family of  states
with  free-energies comparable to the original one? 
The argument against an entropic drive  goes as follows:  While a system in a given phase is  nucleating another  phase $A_o$, it needs to follow a certain  number of steps. Why should this process be accelerated by 
the fact that there where  many options of phases $A_1$, $A_2$ ...,  given that these themselves  require different sets of steps to nucleate?
To clarify the point, let consider a simple situation - a system at very low temperature, activating its escape out of the 
spherical crater $V(r)$ {{of dimension $d$.}}  in Fig. \ref{mexican}. 
Starting  at the bottom, the particle follows (say) a Langevin process, and the probability evolves  via a Fokker-Planck
equation.
\begin{equation}
\dot P= \nabla \left[ T \nabla + \nabla V \right]P
\label{aaaa}
\end{equation}

The probability  distribution was spherically symmetric at the start, and it will remain so, so we may go to spherical coordinates:
\begin{equation}
\dot P= \frac{1}{r^{d-1}} \left[ T \frac{d}{dr} \left( r^{d-1}\frac{dP}{dr} \right)   + P  \frac{d}{dr} \left( r^{d-1}\frac{dV}{dr} \right)  +  \frac{dV}{dr}\frac{dP}{dr}\right]
\end{equation}
Putting ${\tilde P} \equiv r^{d-1}P$ we get the radial diffusion equation
\begin{equation}
\dot {\tilde P}= \left[ T \frac{d^2}{dr^2}    +  \frac{d}{dr}\left(  V(r)-T(d-1) \ln r \right) \right] \tilde P
\label{bbbb}
\end{equation}
{{Thus, the system evolves in an effective potential $V(r) - T(d-1) \ln r$.  The second term in this effective potential derives from the multiplicity of excited states: $(d-1) \ln r$ is the entropy of excited states, being the logarithm of the volume of a shell of radius $r$.  We thus conclude that the multiplicity of excited states have the effect of lowering the barrier against nucleation.}}  
In particular,  with very small $T$, but with {{$T\,d=O(1)$}}   such  that $V(r)-T(d-1) \ln r$ is monotonic decreasing, the particle will escape in finite time.

{{An intuitive way of understanding this result is as follows:}}  At low temperatures, for every trajectory that escapes, there are many
unsuccessful trials which fall just short of escape, and return to the center. Although still extremely rare, trials reaching, say, $90\%$ of the barrier height are exponentially more common
than those which actually escape.  Their frequency  is, however, proportional to the surface of the $90\%$  energy level. 
The situation for one out of an exponential number of  metastable states is very similar. The nucleation of a droplet  of
of any other state of volume  $V$ will cost a free energy at most proportional to the surface.  However, there are $e^{V \Sigma(f)}$ options
of different droplets of free energy density $f$, and they will be accessed with total frequency $\int df \; e^{V [\Sigma-\beta f]}$. These need not be grown simultaneously, but the many unsuccessful attempts in which a droplet grows and collapses
will be   this number of   times  as frequent as if there were only one other option. 
The existence of many paths (in the particle language) or metastable states (in the system language) does not make it easier to nucleate {\it into} any {\it particular} state, but it makes it easier to nucleate {\it out} of the initial deep well.
In other words, if we lump the ensemble of other states  as
 as a single entity (the liquid state, in fact), the probability of  nucleating one of this ensemble (that is, of escaping a specific state)  is just as if
 there were an extra   term in the free energy of the liquid  equal to $-T \ln \left[  \int df \; e^{V \Sigma(f) -\beta f}\right]$.
 Therefore, the usual nucleation argument tells us that this volume term will sooner or later  overcome the surface term,
and the system will leave the original state: this competition leads to escape only if there is an extensive complexity.
 
 Going back to the example of escaping a crater in large dimensions (such that $T\,d$ is of order one), we may describe this at low temperatures
 either by an ensemble of extremal trajectories with different angles (the low temperature treatment  of (\ref{aaaa}), or by a single escape trajectory for the radius (i.e. the low temperature limit as applied to
  Eq. (\ref{bbbb})). The ensemble of  escape routes at different angles in the original space, and the single escape trajectory in radial
 coordinates play  respectively the role, played  in  an extended system 
  by the nucleation of many phases and the nucleation of the 'liquid phase' that encompasses all of them.
   
 Technically, activation out of a state may be calculated by  considering the extremal trajectory, starting in the state and moving away from it by nucleating
 another phase.  The logarithm of the escape time is given by the action of the trajectory.
 In ordinary nucleation, the trajectory goes through a free energy maximum at the  critical droplet radius, and the total action of the escape trajectory is
 of the order of the free energy barrier at the critical droplet.
  The situation here is different, because we have an exponential number of different trajectories moving out of the state, {\it none of which needs to have a finite
critical  droplet radius}. When we sum over all trajectories,  it is the multiplicity of trajectories that decreases the action, in a manner entirely analogous to the way that multiplicity of states   
decreases the total free energy (cf Eq (\ref{froze})).  That is to say, we have a {\it trajectory complexity}
 that decreases the values of the dynamic action, and hence the  logarithm of the escape time.
 If the action of an individual nucleation keeps growing with the surface of the droplet size, but the number of trajectories is proportional
 to the exponential of the droplet volume, then escaping will sooner or later win -- {\it even without a critical radius for an individual passage.}

We thus conclude that the state will be destabilized by the presence of exponentially many others, even if the passage to any individual one is
extremely improbable. 
Note that this argument also implies that it is impossible to have an exponential number of states with super-Arrhenius lifetimes all the way down to zero
temperature, because the Arrhenius escape probability will sooner or later dominate.

The argument we outlined above is close  in spirit to the mathematical results in \cite{Montanari} and to the construction in Ref. \cite{Biroli_Bouchaud,
Kob_Parisi,Cavagna_Grigera}
related to the point-to-set correlation function.
What has been  proven there, at least for lattice systems with discrete variables, is the following: if one considers a closed volume, 
and asks if a  fixed configuration on the surface is correlated with the one in the center (a point-to-set correlation), the maximal radius $r_{max}$
at which this happens diverges if there is a divergent timescale.
Translated into density profiles, this might seem to  suggest that, for $r_{max} \rightarrow \infty$  the surface configuration determines the time-averaged configuration inside,  and this would imply that the complexity is necessarily subextensive, since the number of different
configurations of the boundary is proportional to the exponential of an {\it area}. There is a problem with this argument, as the simple example of classical mechanics
shows. In that case, giving the configuration at two times determines fully a trajectory in between, and yet, the number of possible trajectories may be
exponential ( the Kolmogorov-Sinai entropy need not be zero).   The reason for this is that, because of the Lyapunov instability, to know the configuration in the whole interval
  one needs to know those of the  boundaries  to greater and greater precision --  the number of digits necessary being proportional to the interval of time considered times the Lyapunov exponent.
In conclusion:  the entropy of the bulk is not necessarily proportional to the boundary surface, {\em even in the case that the bulk configuration is  fully determined by the boundary one}, but may be -- in the case of exponential sensitivity to boundary conditions -- also proportional to the  enclosed volume. 

\section{The Kauzmann/ Random First Order scenario}

 Assume that there is a temperature at which there is an equilibrium situation with  permanent density modulations. We now ask: Are there other excited states with infinite lifetimes,  that we may define  in a time-averaged way?
 For example, a ferromagnet  with periodic boundary conditions may be composed of two domains of positive and negative magnetization, separated by an essentially flat (at sufficiently high dimensions)  interface.  Another example is a crystal with  a single straight dislocation. A third example was already mentioned above: in quasicrystals one may have isolated defects,
 the most famous being the `decapod' defect in a Penrose tiling. In addition, in any  non-periodic system, new states may be obtained from translations,
 if this does not violate the boundary conditions.
 We call these {\it states}  because we are considering the ensemble of configurations visited in infinite times (the thermodynamic limit has been taken first) by a system   that started with any one configuration with a given permanent  defect. In other words, we are studying the problem modulo short-lived (such as particle-vacancy) excitations. 
 
  Next, we ask about the free-energy of these states. In many cases, such as the macroscopic dislocations or domain walls mentioned above, their free-energy is infinitely (though 
  non-extensively in the system's volume ${\cal V}$) higher than the ground state. Their contribution to the equilibrium distribution hence vanishes.
In other cases,  it could conceivable happen, and the Wang-tile example cited above is  an example, that an excited state of this kind has a free-energy difference of order one with respect  to the ground state. In that case, the conclusion is that it will have a finite probability in equilibrium, and we are forced to conclude that the equilibrium distribution
 contains several `ergodic components', mutually inaccessible (for a macroscopic system) in finite times.
 Note that the state with a single `serious' defect will typically have many short-lived activated `benign' defects.
  
 We now ask about the number of these states,   in terms of their free energy, at a given temperature.
 If we count how many of these states there are, regardless of their probability of appearing in an equilibrium sample, the logarithm of this number  is the complexity $\Sigma(T)$  (only weakly dependent on $T$).
 The probability  of each state is proportional to $e^{-\beta {\cal{V}} (f_\alpha+  f_o)}$, for a sample of volume ${\cal{V}}$ . 
 The definition of $\Sigma$ is extremely tricky: as soon as $\Sigma(f)$ is of order one, the simple `entropic drive' argument above says that the states cannot be stable. 
 This is most clearly seen in the example of the  Wang-tile system: as soon as we have a finite density of defects, and hence an extensive $\Sigma$, they are  separated by finite distances  and may be annihilated with local rearrangements.
{\it  Strictly speaking, then, $\Sigma$ is only well defined when it is subextensive,  and it otherwise counts 
 states whose definition is timescale-dependent. }
 
\begin{figure}[htbp]
\includegraphics[width=4.5cm]{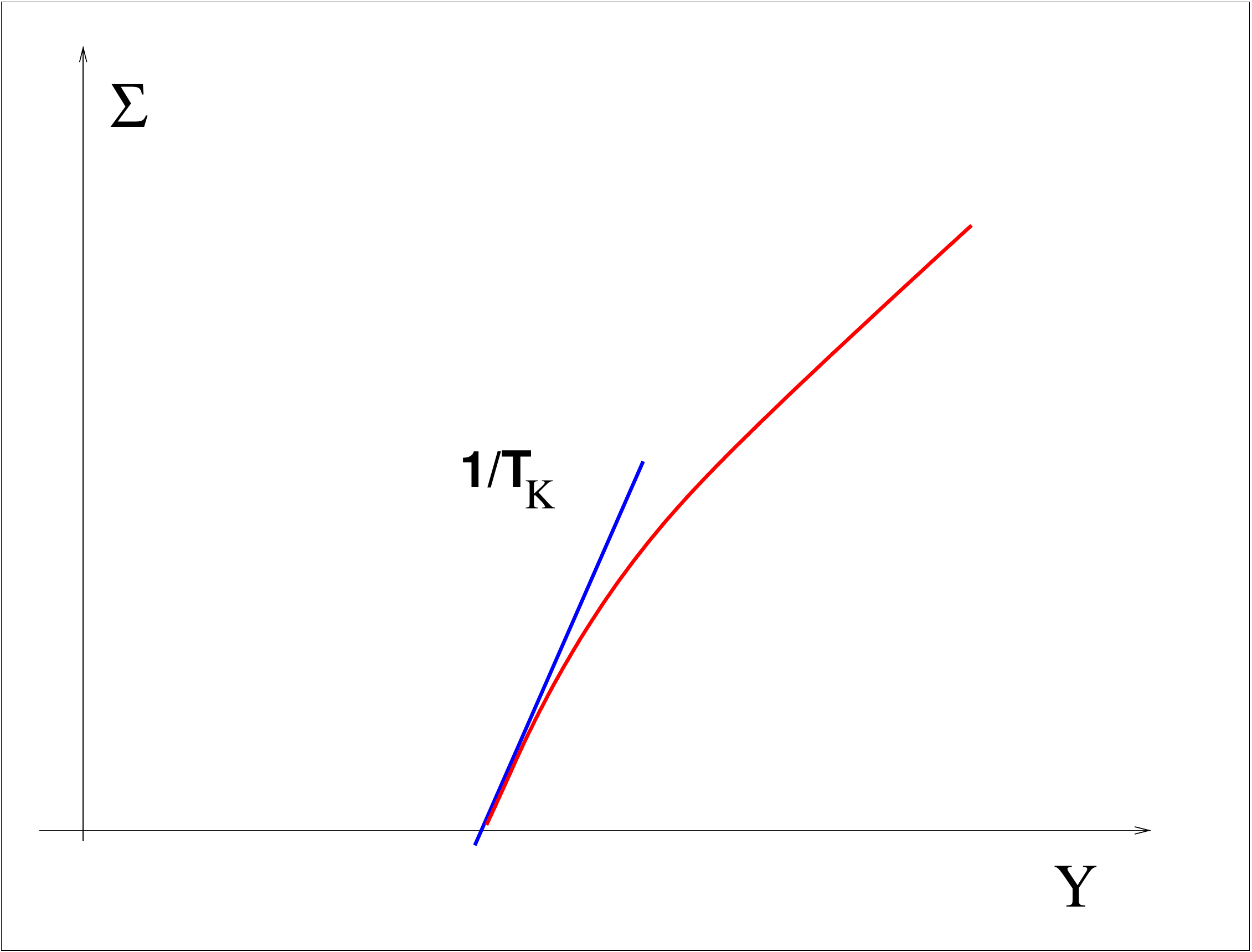}\includegraphics[width=4.5cm]{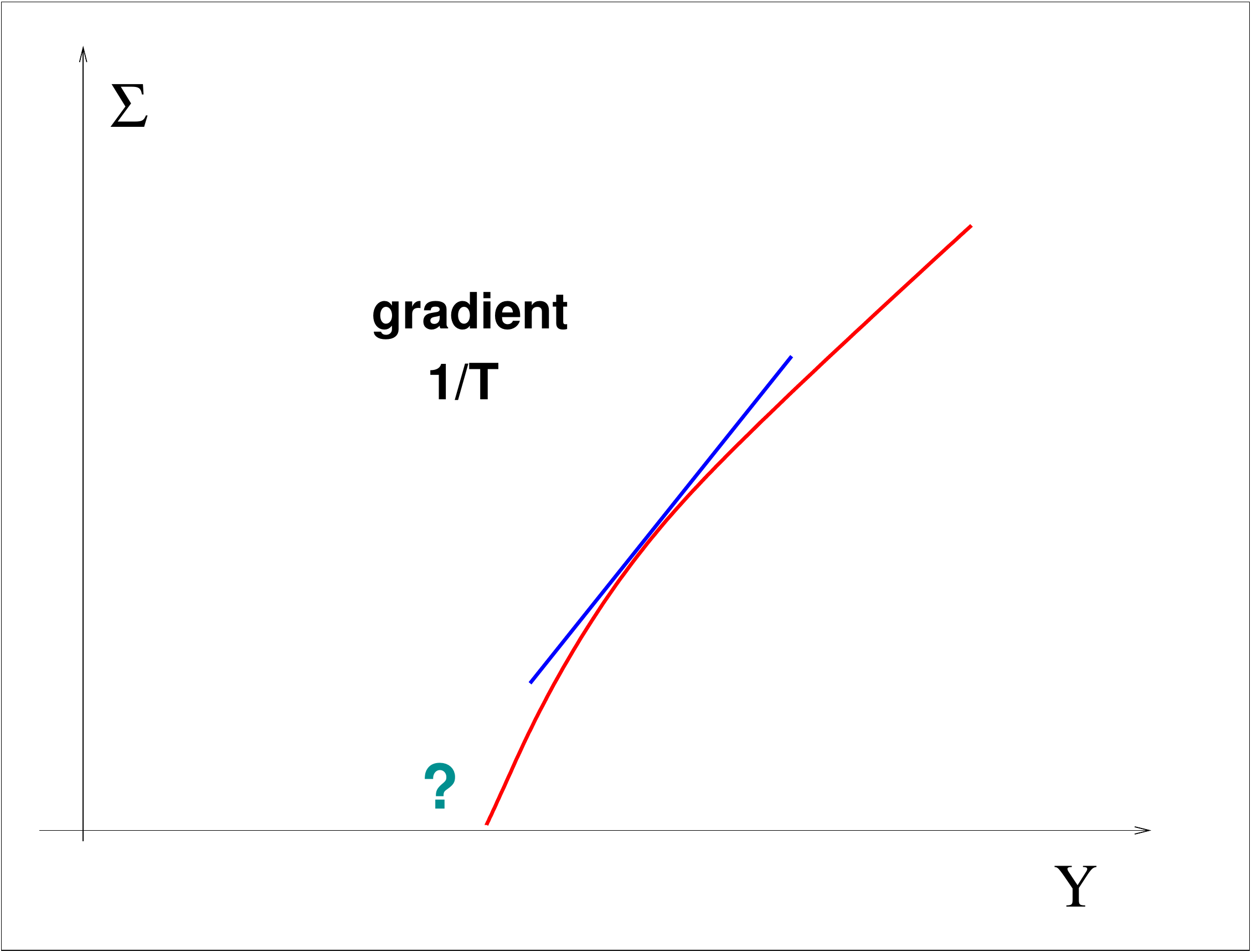}  \includegraphics[width=4.5cm]{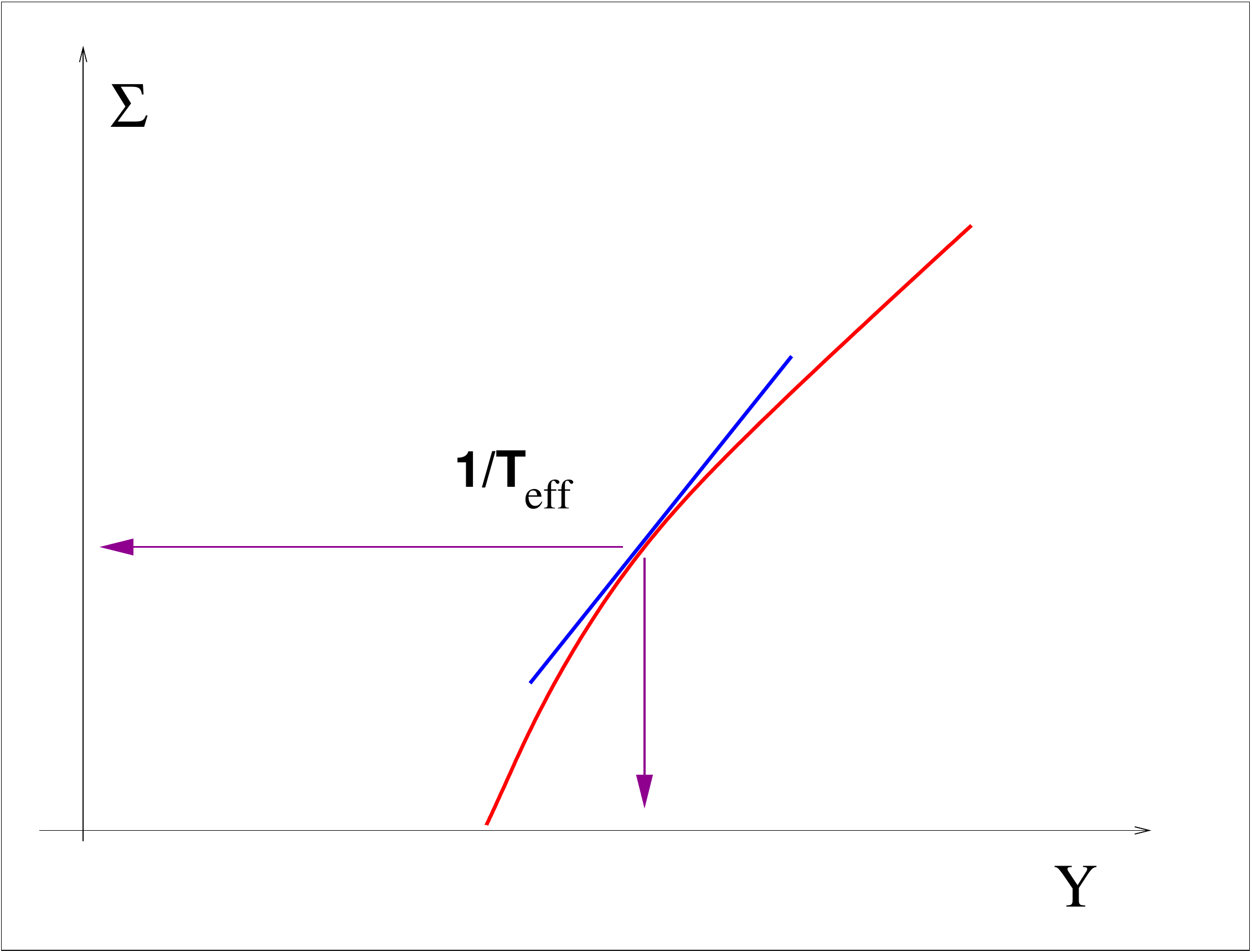} 
\caption{ Complexity of states versus free energy density $Y=f-{\bar f} $. Three situations within the Kauzmann-RFOT scenario. At small temperatures, the system is frozen in the lowest states (left panel).
As soon as $T>T_K$ an exponential number of metastable states dominate, their lifetime is necessarily large but finite, and the system becomes a liquid
(center panel).
Out of equilibrium, the system finds itself exploring states of  free energy higher than those dominating equilibrium. If it does so in essentially a {\it flat} manner, we may assimilate the situation to a two-temperature scenario: $T$ inside a state, and $T_{eff}>T$ between states (right panel)  }
\label{kauzmann}
\vspace{.5cm}
\end{figure}

 As mentioned above (cf. eq. (\ref{froze})), the whole question is whether  $\left.\frac{d\Sigma}{df}\right|_{f=0}< \frac1T $ or  $\left.\frac{d\Sigma}{df}\right|_{f=0}> \frac1T$.  Only in the former case will a distribution  frozen in the lower states be dominant. In the latter case entropy wins,  states with higher
 free energy density dominate, and since these are exponential in number, they are necessarily unstable:  we are in the liquid phase.
 Suppose then that $\left.\frac{d\Sigma}{df}\right|_{f=0}< \frac1T$ holds, and we have a solid with permanent density modulations. As the temperature is increased,  the system eventually becomes a liquid. How is the transition made?
 There are several possibilities:
 
 \begin{itemize}
 \item Second order: the density modulations diminish with increasing temperature, and eventually vanish completely at a temperature $T_*$. This means that 
 the level $q_{EA}$ of the plateau in the autocorrelation decreases continuously and vanishes at a given temperature $T_*$. This is what would
 happen in a ferromagnet, with the plateau value being the magnetization squared.
\item First order: a set of exponentially many (i.e. necessarily unstable) states suddenly begins to dominate the Gibbs distribution above a temperature $T_m$. Such a transition
is like melting. Note that if $\Sigma$ suddenly becomes large, the states will necessarily be individually short-lived, and
the system will jump to a liquid without high viscosity.
\item The $\Sigma$ vs. $f$ curve reaches zero with a finite derivative $\beta_K$ (Figure \ref{kauzmann} ). Because, as mentioned above, 
$\Sigma$ is in general weakly dependent upon temperature, we must state that this has to happen at temperature $T_K$. 
At $T> T_K$ the saddle point  $\left.\frac{d\Sigma}{df}\right|_{f=0}= \frac1T$
begins to dominate. The saddle point complexity starts from zero and grows gradually, unlike the case of melting.
Just above $T_K$, the states that dominate have small complexity and are, because of this, just barely unstable,  the 'entropic pressure' being  very small.
This is the Kauzmann scenario, as revisited from the Random First Order perspective.
\item A particular case of the above is when $\left.\frac{d\Sigma}{df}\right|_{f=0}=\infty$ as $T \rightarrow 0$. In that case, the system has super-Arrhenius 
timescale dependence, but only amorphous order at $T=0$.
\end{itemize}

\begin{figure}[htbp]
\includegraphics[width=6cm]{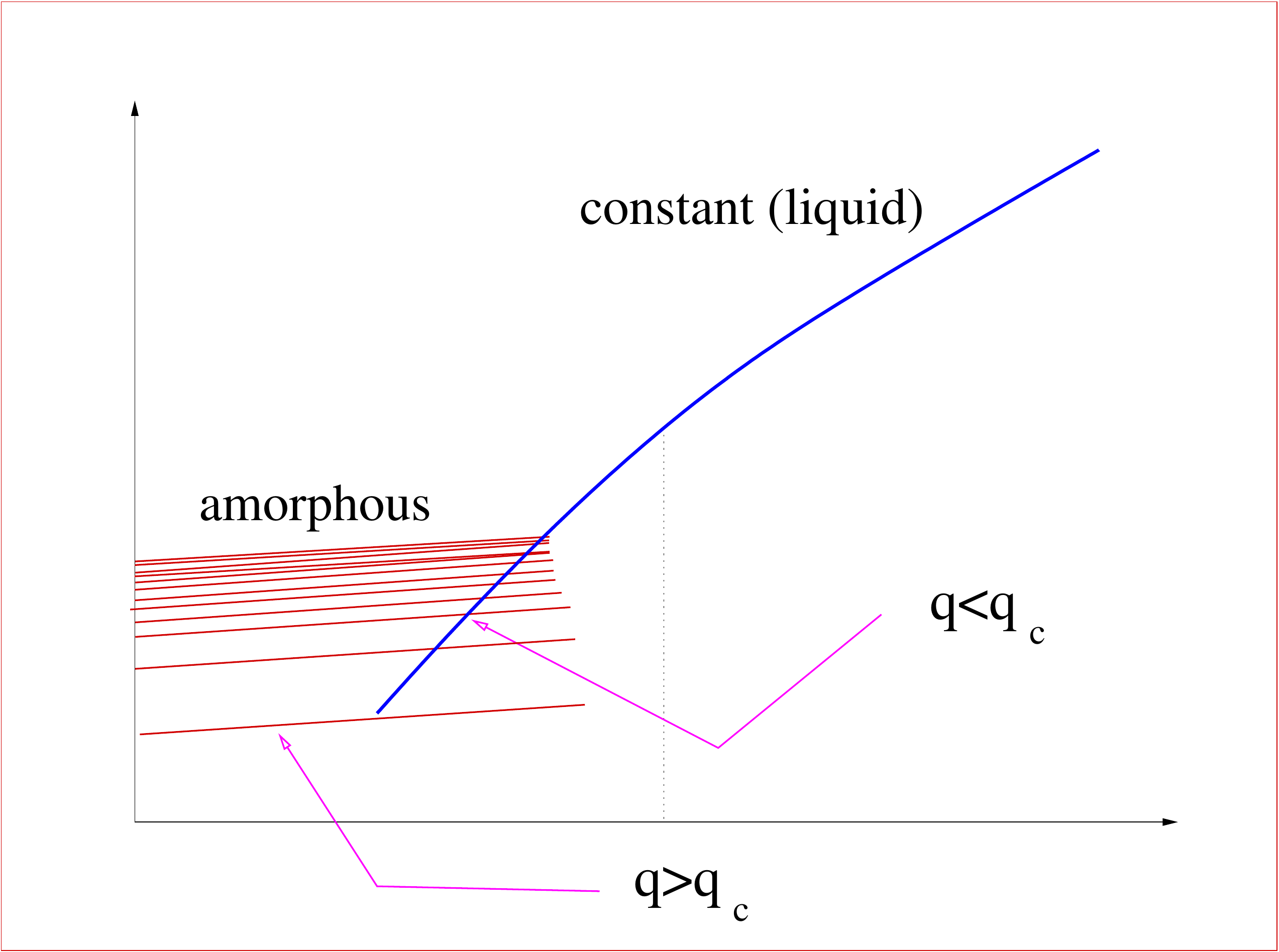}
\caption{ The Renyi entropies of an ideal glass  in equilibrium would be dominated by the lowest states ($q>q_c$), or the higher liquid-like states
$q<q_c$, with $q_c= x$ the Parisi parameter. For the supercooled liquid the situation is the same, except that $q_c>1$, so that typical states are liquid-like.  }
\label{Renyi1}
\vspace{.5cm}
\end{figure}

\subsection{A crucial hypothesis: from global to patch quantities.}

A standard  result  in statistical mechanics  is that the entropy of a  system is calculable, to leading order in the volume, 
as the limit of the patch (or `block') entropy . This is because, even though the boundary conditions of each block are not those of
the total system, one assumes that for short-range correlated systems the effect is of the order of the boundary surface, and hence 
subdominant in the large volume limit.
If we assume that the same holds for the complexity, and furthermore,  for all the Renyi complexities of finite $q$, then we may translate
the predictions of theoretical  scenarios for the entire probability distributions to observable statements concerning the distribution for patches.
In other words, we are led to the assumption that the large-deviation function $G= \frac{1}{V} \ln {\cal P} $, expressed in terms of the normalized logarithmic frequencies $\frac{\ln n}{V}$ , is
the same, to leading order in $V$, as the one obtained for the whole system with volume ${\cal {V}}$, and given boundary conditions.
Note that this  does not mean that {\it individual} profiles dominating patches of given volume $V$ are the same as the state that would dominate
a single patch with, say, periodic boundary conditions.

The Random First Order scenario is strictly speaking, at least for the moment, a mean-field notion, and any extrapolation to finite dimensions is
made at the price of phenomenological thinking (see, however, Ref.\cite{Giulio}). The argument above allows us to make the reasonable assumption that the predictions
of RFOT for the number of states of the entire system may be applied to patches embedded in an infinite system as well.
The RFOT theory predicts, at least for mean-field models, a series of metastable states with internal free-energy $f_\alpha$. For a given $f$, there 
is a quantity ${\cal N}(f)$ of these states, with $\ln {\cal{N}} \sim {\cal{V}} \Sigma^{sys}(f) $, which defines the complexity $\Sigma^{sys}$ --  a function that
depends only weakly on temperature, because states retain their integrity  when temperature changes.
The  assumption that all Renyi complexities of the whole system may be computed as a limit of the Renyi complexities for large patches means that
we may identify $\Sigma^{sys}$ and $\Sigma$. Thus:
\begin{eqnarray}
p^{(equil)} &=&  \frac{e^{-V \beta  f_\alpha}}{\sum_{\alpha'  } e^{-V \beta  f_\alpha'}}
\nonumber \\
K_q &\sim & \frac{1}{  \; V (q-1) } \left\{\ln \left[\sum_\alpha    e^{-V  q \; f_\alpha}\right] -  q \ln \left[\sum_\alpha    e^{-V   \beta \; f_\alpha}\right] \right\} \nonumber \\
&= &   \frac{1}{ \; V (q-1) }\left\{  \ln \int df  \; e^{ \; V [ \Sigma(f) - \beta q f]  } - q\ln \int df  \; e^{ \; V [ \Sigma(f) - \beta  f]  } \right\}\nonumber \\
&=& \frac{1}{ \; V (q-1) }  \ln \int df  \; e^{ \; V [ \Sigma(f) - \beta q (f-{\bar f})]   } 
\label{close}
\end{eqnarray}
where 
\begin{equation}
\beta {\bar f} \equiv  - \frac{1}{V} \ln \int df  \; e^{ \; V [ \Sigma(f) - \beta  f]    } 
\end{equation}
Comparison with equation ({\ref{iii}) yields, {\it and we recognize $q$ as the $m$ parameter of replica theory}~\cite{MePaVi}. Varying the Renyi parameter 
$q$ is then 
exactly the procedure introduced by Monasson~\cite{Remi} to analyze these models (see Fig. \ref{Renyi1}). 
For example, this means that if the system were a one-step replica symmetry breaking solution in the low temperature phase, with Parisi parameter
$x$,  then the Renyi complexities would be (see Fig. \ref{Renyi}):
\begin{eqnarray}
K_q&=&0 \;\;\;\;\; for \;\;\;  \;\; q<x \nonumber \\
K_q&>&0 \;\;\;\;\; for \;\;\;  \;\; q>x
\label{renyi1}
\end{eqnarray}
{{This is}} directly measurable if an equilibrium  configuration is available (Fig \ref{Renyi}).

Experimentally, we thus have to proceed as follows. We first define an averaged profile, over a time $\tau_\rho$ of the order of $\tau_\alpha$. 
{\it Note that $\tau_\alpha$ is not the same for all patches, with the rarer ones expected to have shorter lifetimes}. Hence, we must be careful 
of the fact that the number of rare states (i.e. the Renyi complexities for large $q$) may be dependent on the averaging time $\tau_\rho$.

Next, we either compute the Renyi complexities \`a la Grassberger-Procaccia, by scaling the results for  different $V$ and different $\epsilon$, or , alternatively,
we set up a large deviation function plotting the logarithm of the probability of having a number of repetitions versus the logarithm of the 
 number of repetitions, both normalized by $V$.
 
 {\it A fingerprint of the RFOT scenario is then a non-zero crossover Renyi parameter $q=x$, implying that there are always rare 
 patches that repeat exponentially {infrequently}}. On approaching the transition from below $x \rightarrow 1$, and this means that  non-repeating  patches become more and more frequent, and eventually typical at $T=T_K$}.
 Conversely, in the liquid phase just above $T_K$, most patches repeat exponentially rarely. However,  there are a few  ones that repeat
 frequently, but their number is so small  that even their high repetition rate does not allow them to dominate the measure.
 
 Let us mention here that we may take the point of view of patches, rather than the whole system, for more complicated (and informative) large-deviation
 functions, depending  on the patch overlap  $Q$, defined, for example  as:
 \begin{equation}
 Q_{ab}= \frac{1}{V} \int d{\bf r} \; (\rho_a({\bf r})-\bar \rho) (\rho_b({\bf r})-\bar \rho)
 \label{QQ}
 \end{equation}
One takes a patch at random, and counts the average complexity $K_1(Q)$  as in Equation  (\ref{GP4}), but restricting the sum 
to patches at distance $Q$. Finally, one averages over initial patches. $K_1(Q)$ is the patch version of the Franz-Parisi effective potential \cite{FrPa},
a function closely related to the Parisi ansatz, for which there are also suggestions  within the RFOT scenario.

 \section{Out of equilibrium: effective temperatures}
 
 In practice, glasses are out of equilibrium, and the concept of temperature does not apply in principle to them.
 An old idea~\cite{Tool}  has been to assume that  the structure somehow ``remembers'' the situation of the liquid at the temperature it had
 before  it fell out of equilibrium. The system would have one (or, in later versions,  several) ``fictive temperatures'' governing the slow evolution.  
 On the other hand, when analytic solutions for the glassy dynamics of mean-field models became available~\cite{CuKu}, it turned out that these systems 
 possessed an effective temperature $T_{eff}$ relating fluctuation and dissipation (of {\it any observable}) at low frequencies \cite{CuKuPe}.
 Indeed, $T_{eff}$ would be measured by applying a low-frequency thermometer to the system out of equilibrium, an experiment that has been performed \cite{Grigera}.
 The interpretation for the appearance of this  temperature is not yet fully clear, but it is {\it not} related to structures that are frozen in the form they had
 when the system crossed the transition, but rather to structures that form again and again. We know this because effective temperatures  exist at times which are long compared to the  autocorrelation time, so the system had more than enough  time to restructure.
  
 A somewhat strong assumption is to suppose that this temperature arises because a system out of equilibrium explores metastable states with
 a  probability at $T_{eff}$  -- a proposal that agrees with a hypothesis that had been made by Edwards for granular matter 
 \cite{Edwards,FV,SEKU} -- and that $T_{eff}$ slowly evolves in a much longer (perhaps infinite) timescale towards the equilibrium value $T$.
 In this very strong form, the assumption is that if the system has metastable states with free energy $f_\alpha$, so that the equilibrium probability
 would be proportional to $e^{- f_\alpha/T}$; out of equilibrium the probabilities would be those associated  with a probability $\propto e^{- f_\alpha/T_{eff}}$.
 In other words, if the probability of a certain configuration in equilibrium is $p_\alpha^{(equil)}$, the probability out of equilibrium should be:
 \begin{equation}
 p_\alpha^{(out \; of \; equil)}  \propto \left[p_\alpha^{(equil)}\right]^{\frac{T}{T_{eff}}} 
\end{equation}

 Again, if we assume that what has been said for ${\cal{V}}$ can also be said for large $V$, then we should have:
\begin{equation}
V(q-1) K_q = \ln \left(\sum_\alpha e^{- q \beta_{eff} f_\alpha}\right) - q  \ln \left(\sum_\alpha e^{-\beta_{eff} f_\alpha}\right)
\end{equation}
which means that, in every physical situation, the dependence on $q$ may be written as
\begin{equation}
\frac{(q-1)}{q}  K_q = \beta_{eff} \; \left[ U(q \beta_{eff}) - U(\beta_{eff}) \right]
\label{scal}
\end{equation}
where $U(b) \equiv \frac{1}{b} \sum_\alpha e^{-b f_\alpha}$ is a  function that depends, upon this assumption,  on the final  parameters (temperature, pressure).
 All the history of the sample would manifest itself, under this hypothesis, only through the single parameter  
 $T_{eff}=1/\beta_{eff}$. In the limit $q \rightarrow 1$ (\ref{scal}) reduces to $K_1=\Sigma|_{\beta_{eff}}$, the complexity evaluated in the point where $\Sigma' = \beta_{eff}$. 

In order to check the hypothesis of an effective temperature of this kind, we should repeat different protocols, and for each determine the value of
$T_{eff}$ that will put the results in the form (\ref{scal}), with  a single protocol-dependent $T_{eff}$ for all $q$'s.  This should be possible {{in principle}}, and if this is so, one could check that the patterns that dominate
the measure at a given $T_{eff}$ are the same, irrespective of the protocol by which it has been reached.

\section{Conclusion}
In this paper we have discussed how to render observable several quantities from glass theory, in particular those that 
have their origin in mean-field like theories. 
We have discussed a correlation length that, we argue, should diverge for any solid, defined as a system in which, in spite of thermal
fluctuations, there are permanent spatial density modulations. The construction has the strength of not being tied to any
specific phenomenological scenario.
Once we  define states, order and lengths in a concrete and experimentally measurable manner, we are lead to the conclusion that
the mere existence of a  timescale that goes to infinity  faster than an Arrhenius law 
implies, for finite, short-range interactions,  that the  lengthscale diverges.  

Defining further quantities associated with patches embedded within the system, we may answer the question of how, on the basis of  the data from an ideal glass configuration, we could 
recognize if it corresponds to a Random First order/ Kauzmann scenario. 
Perhaps more relevant for realistic systems, we may also use the technique to decide, using the experimental   out of equilibrium data, 
 whether there is a flat exploration of metastable states, leading to an effective temperature.

\begin{acknowledgments}

We wish to thank S. Aubry, L. Berthier, G. Biroli, S. Franz, Y. Kafri, N. Merhav, R. Schilling  and F. Zamponi for illuminating discussions. DL 
gratefully acknowledges support from the Israel Science Foundation, under Grant 1574/08.

\end{acknowledgments}


\begin{thebibliography}{100}
\bibitem{foot00}In the case of crystals, we have the alternative of spatial, rather than temporal, averaging, which is 
automatically done by Fourier transform. We do not know how to do this passage from time to space averaging in  non-periodic systems.
\bibitem{foot22} {Note that to obtain a continuous $\rho$ we have already smoothened by a much smaller scale Equation (\ref{smo})}
\end{thebibliography}
\end{document}